\documentclass[longauth]{aa} 

\usepackage {epsfig,graphicx,color}
\usepackage {txfonts   }
\usepackage {natbib    }
\usepackage {float     }
\usepackage {enumerate }
\usepackage {tabularx  }
\usepackage {url       }
\usepackage {capt-of   }
\usepackage {color     }

\begin{document}

%

\title{Ionization compression impact on dense gas distribution and star formation}
\subtitle{Probability density functions around \ion{H}{ii} regions as
  seen by {\it Herschel}\thanks{Herschel is an ESA space
    observatory with science instruments provided by European-led
    Principal Investigator consortia and with important participation
    from NASA.}}

\author{ 
P. Tremblin    \inst{1,2}\and 
N. Schneider   \inst{1,3,4}\and 
V. Minier      \inst{1}\and
P. Didelon     \inst{1}\and
T. Hill        \inst{1,5}\and
L. D. Anderson \inst{6}\and
F. Motte       \inst{1}\and
A. Zavagno     \inst{7}\and
Ph. Andr\'e    \inst{1}\and
D. Arzoumanian \inst{8}\and
E. Audit       \inst{1,9}\and
M. Benedettini \inst{10}\and
S. Bontemps    \inst{3,4}\and
T. Csengeri    \inst{11}\and
J. Di Francesco\inst{12}\and
T. Giannini    \inst{13}\and
M. Hennemann   \inst{1}\and
Q. Nguyen Luong\inst{14}   \and  
A. P. Marston   \inst{15}\and
N. Peretto     \inst{16} \and
A. Rivera-Ingraham\inst{17,18}\and
D. Russeil      \inst{7}\and
K. L. J. Rygl   \inst{10} \and
L. Spinoglio    \inst{10} \and
G. J. White     \inst{19,20}
       }

\institute{Laboratoire AIM Paris-Saclay (CEA/Irfu - Uni. Paris Diderot
  - CNRS/INSU), Centre d'\'etudes de Saclay,  91191 Gif-Sur-Yvette,
  France
  \and
  Astrophysics Group, University of Exeter, EX4 4QL Exeter, UK
  \and
  Univ. Bordeaux, LAB, UMR 5804, F-33270, Floirac, France
  \and 
  CNRS, LAB, UMR 5804, F-33270, Floirac, France
  \and  
  Joint ALMA Observatory, Alonso de Cordova 3107, Vitacura, Santiago, Chile
  \and
  Department of Physics, West Virginia University, Morgantown, WV 26506, USA
  \and
  Aix Marseille Universit\'e, CNRS, LAM (Laboratoire d'Astrophysique de
  Marseille) UMR 7326, 13388, Marseille, France  
  \and
  IAS, CNRS (UMR 8617), Universit\'e Paris-Sud, B\^atiment 121, 91400
  Orsay, France
  \and
  Maison de la Simulation, CEA-CNRS-INRIA-UPS-UVSQ, USR 3441, Centre
  d'\'etude de Saclay, 91191 Gif-Sur-Yvette, France
  \and
  Istituto di Astrofisica e Planetologia Spaziali (INAF-IAPS), via del
  Fosso del Cavaliere 100, 00133 Roma, Italy 
  \and
  Max-Planck Institut f\"ur Radioastronomie, Auf dem H\"ugel,
  Bonn, Germany
  \and 
  National Research Council of Canada, Herzberg Institute of
  Astrophysics, 5071 West Saanich Road, Victoria, BC V9E 2E7, Canada 
  \and
  INAF Osservatorio Astronomico di Roma, via Frascati 33, 00040 Monte
  Porzio Catone, Italy 
  \and
  Canadian Institute for Theoretical Astrophysics, University of
  Toronto, 60 St. George Street, Toronto, ON M5S 3H8, Canada 
  \and
  European Space Astronomy Centre, Urb. Villafranca del Castillo, PO
  Box 50727, E-28080 Madrid, Spain 
  \and
  School of Physics and Astronomy, Cardiff University, Queens
  Buildings, The Parade, Cardiff CF24 3AA, UK 
  \and
  Universit\'e de Toulouse; UPS-OMP; IRAP;  Toulouse, France
  \and
  CNRS; IRAP; 9 Av. colonel Roche, BP 44346, F-31028 Toulouse cedex 4, France
  \and
  The Rutherford Appleton Laboratory, Chilton, Didcot, OX11 0NL, UK
  \and
  Department of Physics and Astronomy, The Open University, Milton Keynes, UK
}

\date{\today}
\mail{pascal.tremblin@cea.fr}

\titlerunning{Multi-component PDFs}
\authorrunning{Tremblin et al.}

\abstract{}
{The expansion of hot ionized gas from an \ion{H}{ii} region into a
  molecular cloud compresses the material and leads to the formation
  of dense continuous layers as well as pillars and globules in the
  interaction zone.  $Herschel$ imaging in the far-infrared (FIR) confirmed the
  presence of these dense features -- potential sites of
  star-formation -- at the edges of \ion{H}{ii} regions. This feedback
  should also impact the probability distribution function (PDF) of
  the column density around the ionized gas. We aim to quantify this
  effect and discuss its potential link to the Core and Initial Mass Function (CMF/IMF).}
{We used in a systematic way $Herschel$ column density maps of several
  regions observed within the
  HOBYS key program:
  M16, the Rosette and Vela C molecular cloud, and
  the RCW 120 \ion{H}{ii} region. We computed the PDFs in concentric
  disks around the main ionizing sources, determined their
  properties, and discuss the effect of ionization pressure on the distribution
  of the column density.}
{We fitted the column density PDFs of all clouds with two lognormal
  distributions, since they present a 'double-peak' or 
enlarged shape in the PDF. Our interpretation is that the
lowest part of the column density distribution describes the
turbulent molecular gas while the second peak corresponds to a
compression zone induced by the expansion of the ionized gas into the
turbulent molecular cloud. Such a double-peak is not visible for all clouds
associated with ionization fronts but depends on the relative
importance of ionization-pressure and turbulent ram pressure. A
power-law tail is present for higher column densities, generally ascribed to the
effect of gravity. 
The condensations at the edge of the ionized gas have a steep
compressed radial
profile, sometimes recognizable in the flattening of the power-law
tail. This could lead to an unambiguous criterion able to disentangle triggered
from pre-existing star formation.} 
{In the context of the gravo-turbulent scenario for the origin of the
 CMF/IMF, the double
 peaked/enlarged shape of the PDF may impact the formation of objects at both the
 low-mass and the high-mass end of the CMF/IMF.
  In particular a broader PDF is required 
  by the gravo-turbulent scenario to fit properly the IMF with a
  reasonable initial Mach number for the molecular cloud. Since other
  physical processes (e.g. the equation of state and the variations
  among the core properties) have already been suggested to broaden
  the PDF, the relative importance of the different effects remains an
open question.}


\maketitle

%
%

\section{Introduction}

\begin{table*}  
\caption{Molecular cloud parameters.} \label{hobys}   
\begin{center}  
\begin{tabular}{lccccccccc}  
\hline \hline   
Cloud & Distance & Res.$^a$ & $\langle N(H_2) \rangle^b$ & Mass$^c$ &
$\langle T \rangle^d$ & $\langle UV-flux \rangle^e$   &
r$_1$, r$_2$, r$_3$, r$_4$ $^f$\\  
         &  [kpc]    & [pc]  & [10$^{22}$ cm$^{-2}$] & [10$^4$
  M$_{\odot}$] & [K] & [G$_\circ$] & [pc] \\ 
\hline       
M16        &     1.8 & 0.35  & 1.23 & 37.0 & 10.7 (6-23)  & 280
(20-1.8 10$^4$) & 3, 5, 10, 15\\  
Rosette    &     1.6 & 0.28  & 0.31 &  9.9 & 23.4 (12-36) & 10
(1-10$^4$) & 20, 25, 39, 54\\  
RCW120     &     1.3 & 0.23  & 0.29 &  0.6 & 17.2 (13-25) & 125
(4-8000) & 2.5, 4.3, 5.5, 6.8\\  
Vela C/RCW36 &     0.7 & 0.12  & 0.75 &  3.1 & 14.7 (11-29) & 230
(50-1.5 10$^4$) & 1.4, 3.2, 5.2, 6.6\\  
\hline   
\end{tabular}  
\end{center}  
\vskip0.1cm  
\noindent $^a$Spatial resolution scale of the column density map (37$''$).  \\  
$^b$Averaged (over the whole map) column density of gas and dust,
assuming a gas to dust ratio of 100. Note that these
values may differ from the average value in the largest disks (regions
1+2+3+4) used for the PDFs.\\ 
$^c$Total mass from column density map above N(H$_2$)=10$^{21}$
cm$^{-2}$ using the conversion 
formula N(H$_2$)/$A_V$=0.94$\times$10$^{21}$ cm$^{-2}$ mag$^{-1}$ \citep{Bohlin:1978dw}. \\ 
$^d$Average dust temperature (total range in parenthesis) from {\sl
  Herschel} data. \\  
$^e$Average and min/max UV-flux (Schneider, priv.comm.) in Habing field, 
determined using the 70 $\mu$m and 160 $\mu$m {\sl Herschel}
flux. Details of the method are described in
\citet{Roccatagliata:2013wv}. For RCW120, the typical value in the PDR-zone is 1-2 10$^3$ G$_\circ$, 
consistent with what determined by Rod\'on et al. (in prep.) using the
spectral type of the exciting star.\\
$^f$ Radii of the different disks used to compute the PDFs.
\end{table*}   

 The role of density compression by the expansion of ionized gas
  into a molecular cloud has been discussed since the pioneering
  theoretical work of \citet{Elmegreen:1977iq}. Dense features are
  observed at the edge of \ion{H}{ii} regions, i.e., condensations
  \citep[e.g.][]{Deharveng:2009kd,Zavagno:2010jv}, globules
  \citep[e.g.][]{Schneider:2012hz}, and pillars
  \citep[e.g.][]{Hester:1996ir,Schneider:2010ec}.  Although many models and numerical
  simulations have been able to explain how small structures can form
  \citep[see][]{Bertoldi:1989bq, Lefloch:1994ts,Miao:2006bx,Miao:2009ds,Mackey:2010cv,
    Bisbas:2011kg, Gritschneder:2010du, Haworth:2011gv,
    Tremblin:2012ej, Tremblin:2012he}, the question remains whether
  they are triggered or pre-existing.  Simulation on the molecular
  cloud scale showed that the impact of high-mass stars on the molecular
  gas can enhance star formation \citep{Dale:2007cr} or that the
  impact is rather negligible \citep[see][]{Dale:2011ct}, depending on
  the physical properties of the cloud.  In \citet{Tremblin:2012he},
we showed that the key parameter to understand the impact of a high-mass
star on a molecular cloud is the ratio of the ionized-gas
pressure to the ram pressure of the turbulence of the cloud. When the
ionized-gas pressure dominates, compression from ionization is
important, whereas the inverse leads to a compression dominated by the
effect of turbulence.  Turbulent compression has
  been intensively studied in the past few years thanks to simulations of
  isothermal supersonic turbulence
  \citep[see][]{VazquezSemadeni:1994fm,Padoan:1997uu,Kritsuk:2007gn,VazquezSemadeni:2008ho,Federrath:2008ey,Federrath:2010ef}. They
  showed that the probability distribution function of the density
  (density PDF) is well approximated by a lognormal form. Deviations
  from the lognormal shape -- mostly in the form of power-law tails --
  were seen in simulations of compressible turbulence
   of a non-isothermal gas distribution
  \citep{Passot:1998cr}, and/or in models including self-gravity
  \citep{Klessen:2000ca,Kritsuk:2011hw,Federrath:2013ip}. The effect of magnetic
  fields on the density PDF was found to be less important, generally
  reducing the standard deviation
  \citep{Molina:2012iv,Federrath:2012em,Federrath:2013ip}. 

 The impact of ionization on a turbulent velocity field was
   studied by \citet{Gritschneder:2010du} and the effect of
 ionization compression on the PDF in turbulent simulations 
  has been investigated only recently by \citet{Tremblin:2012he}. They
  showed that when the ionized-gas pressure is larger than the ram
  pressure of the turbulence, a second peak in the PDF forms at high
  densities in addition to the lognormal shape. 
 Observationally, this second peak caused by the compression from
  the ionized-gas pressure has been identified in the
  Rosette Nebula by \citet{Schneider:2012ds}.  But thanks to recent
  FIR-observations with the  {\it Herschel} space telescope, it is
now possible to study the column density PDF of the cold and warm 
molecular gas in high-mass star-forming regions, in which 
compression from ionization is likely to take place. Such regions are
observed within the
HOBYS\footnote{\url{http://hobys-herschel.cea.fr}} key program
\citep[see][]{Motte:2010fy,Motte:2012ut} and 
our study focuses on  typical examples, i.e. the Rosette
\citep{Motte:2010fy,Schneider:2010ec} and M16 
molecular clouds \citep{Hill:2012jb},  and the  \ion{H}{ii} regions
RCW 120 \citep{Zavagno:2010jv,Anderson:2010dp,Anderson:2012jy} and RCW 36 
\citep[in the Vela C molecular cloud, ][]{Hill:2011ht,Minier:2013ih}.

We determined the PDFs in areas of four concentric disks around the
\ion{H}{ii} regions present in these clouds in order to study the
large-scale effect of the ionization on the cold molecular gas. The
first disk always contains the OB cluster region and includes the closest dense
structures of the molecular cloud. The centre of the disk is not the
main ionizing source but corresponds to the approximate centre of the
H$\alpha$ emission that is also approximately the centre of the
spherical column density structures seen at the edge of the cavity in
the {\it Herschel} maps. For Rosette and M16, the other disks
are enlarged depending of the various dense structures present in the
column density maps (37$''$ angular resolution), up to the largest
disks that contain most of the maps.  This choice allows us to
  study the evolution of the shape of the PDF in function of the
  column density structures that are included in the disks. For RCW
  120 and RCW 36, less dense structures can be identified and the different radius are
  regularly spaced between the inner disk and the largest one reaching
the edge of the observed map. Overall, the parameters of the fits of
the PDFs have only little 
dependence on the precise choice of the radii of the disks (given in
Table~1) as long as they include the same dense column density
  structures that can be seen in the {\it Herschel} maps (see an
  example for Rosette in Appendix~A). 
We first present and validate the method on the M16 molecular cloud
(Sect.~\ref{sect_m16}), before showing the results on the Rosette
Nebula (Sect.~\ref{sect_rosette}). Both of these regions are relatively
large and their \ion{H}{ii} region extended ($\approx$ 10-20 pc).
 However, their mass and the incident UV-flux is rather different (see
  Table~\ref{hobys}). Finally, we study smaller-scale \ion{H}{ii}
regions, RCW 120 and RCW 36 (Sect.~\ref{sect_rcw120} and
\ref{sect_rcw36}), that are only $\approx$1\,pc large and have
  lower masses. This approach allows us to explore the
  compression effect over one order of magnitude each of spatial scale,
  mass, and incident UV-flux.

\section{M16}\label{sect_m16}

M16 is located in the constellation of Serpens at 1.8 kpc
from the Sun \citep{Bonatto:2006kb}. The young stellar cluster NGC
6611 is ionizing the molecular gas of this star forming region. The
principal ionizing sources are one O4 and one O5 stars whose combined
ionizing flux is of order $2\times 10^{50}$ s$^{-1}$
\citep{Hester:1996ir,White:1999ue}. The discovery of pillars of gas in
M16 by
the Hubble space Telescope \citep{Hester:1996ir} popularized this
region  by naming it ``Pillars of Creation''.  Many spectral line
studies have been performed on the cloud \citep[see][among
others]{Pound:1998hj,White:1999ue,Allen:1999cl,Urquhart:2003jj}.
 The velocity of the CO line emission for the pillars is found between $\sim$20 and
  $\sim$30 km s$^{-1}$ \citep{White:1999ue}. These authors
  suggest that emission around 29 km s$^{-1}$ is probably unrelated to
  the pillars. However with regard to the vicinity of the velocity to the
  cloud's bulk emission and its spatial correlation (see their
  Fig.~8), a physical relation with the rest of the cloud is very
  likely. We thus do not 
  expect the {\sl Herschel} column density map to be strongly affected
  by line-of-sight (LOS) confusion.
Recently, \citet{Flagey:2011jr} measured dust spectral energy
distributions (SEDs) thanks to {\it Spitzer} data. They show that the
SED cannot be accounted for by interstellar dust heating by UV
radiation but an  additional source of radiation is needed to match
the mid-IR flux. They proposed two explanations for this source of
pressure: stellar winds or a supernova remnant. 
{\it Herschel} provides new observations on the region and
\citet{Hill:2012jb} studied in detail the impact of the ionizing
source on the temperature of the molecular gas. Heating is observed in
regions with $n_{H_2} \approx 10^5-10^6$ cm$^{-3}$ thus impacting the
initial conditions of the star-forming sites. We use the same {\it
  Herschel} map to study the PDF of the cold molecular gas at the
interface with the \ion{H}{ii} region. The column density map of the
M16 molecular cloud (and of the other regions) was made by fitting
pixel-by-pixel the spectral energy 
distribution (SED) of a greybody to the Herschel wavebands between 160
and 500 $\mu$m (at the same 37$''$ resolution), assuming the dust opacity
law of \citet{Hildebrand:1983tm} and a spectral index $\beta$ of 2.  

\begin{figure}[t]
\centering
\includegraphics[trim=0 3cm 2cm 0,width=\linewidth]{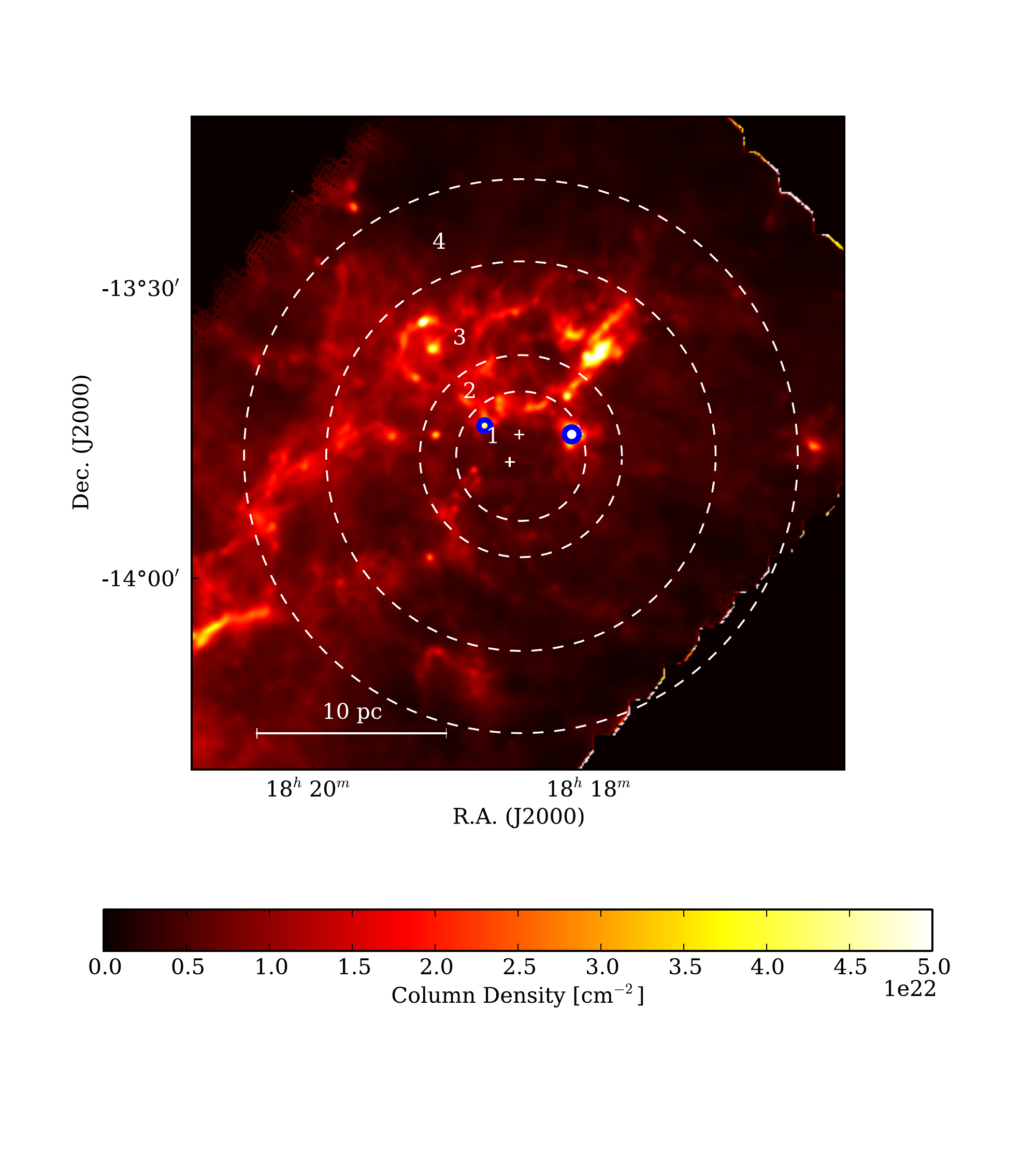}
\caption{\label{M16_N_pdf} M16 {\it Herschel} column density map
  \citep{Hill:2012jb}. The circles indicate the different regions used
  for the PDFs. The white crosses indicate the position of the main ionizing
  sources. The limit of the H$\alpha$
  emission is inside region 1+2+3 showing approximatively the limit of the \ion{H}{ii} region (although
  the \ion{H}{ii} region is not spherical in M16). The blue circles
  indicate the position of the dense condensations 
  identified in the power-law tail of the PDF of region 1.}
\end{figure}

\begin{figure}[t]
\centering
\includegraphics[trim=0 0 1cm 0,width=\linewidth]{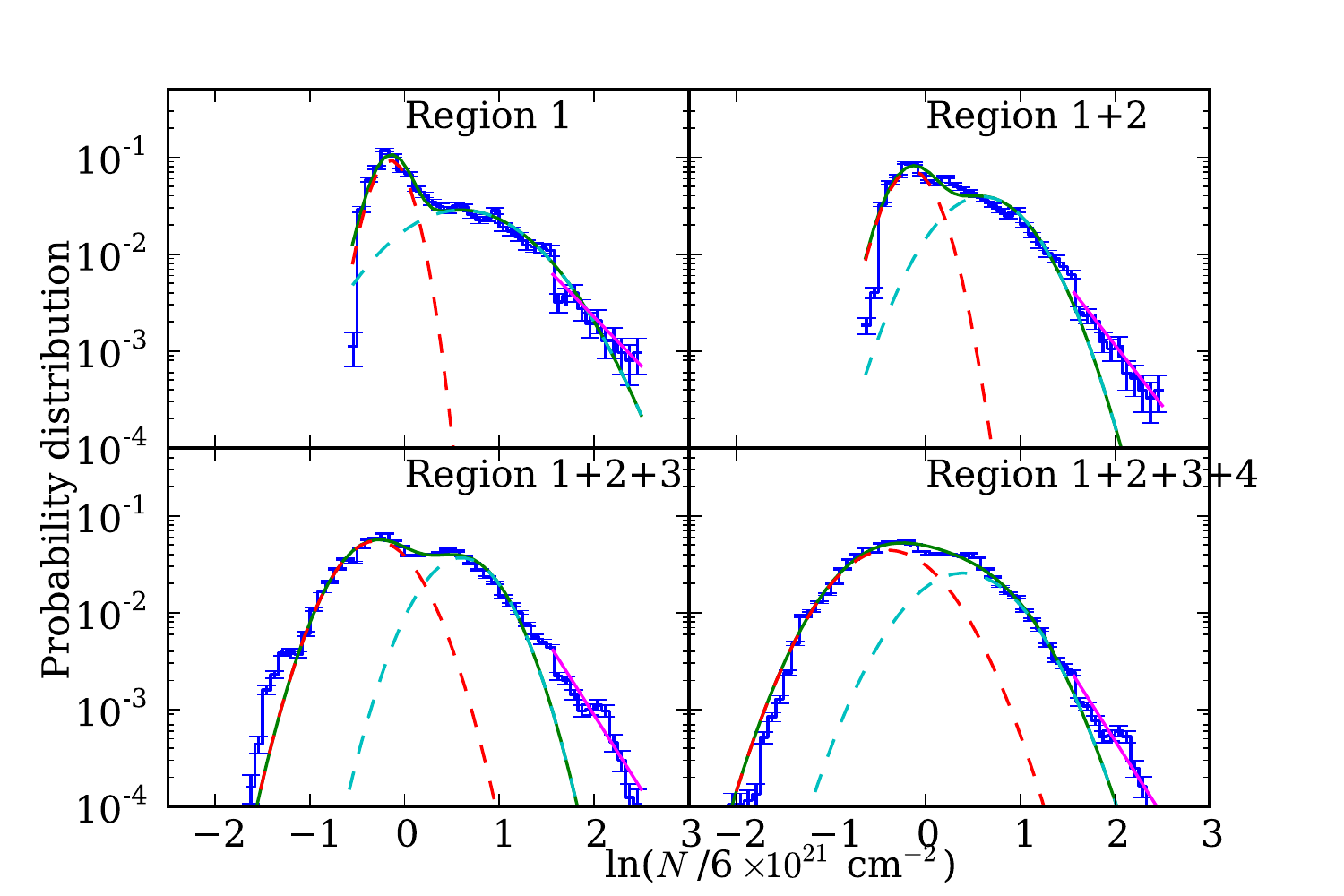}
\captionof{figure}{\label{M16_pdf} M16 PDFs on the four regions indicated
  in Fig. \ref{M16_N_pdf}. The multi-component fit (green curve) is done using
  the sum of two lognormal distributions (red and blue dashed curves, see Eq. \ref{eq_fit}) and a power law at
  high column densities (magenta line, see Eq. \ref{eq_pl}). The
  error-bars are computed assuming Poisson noise in each bin of the
  distribution.}

\textcolor{white}{.}\\
\begin{tabular}{l|ccccccc}
Region & $\eta_0$ & $p_0$ & $\sigma_0$ & $\eta_1$ & $p_1$ & $\sigma_1$ & $m$\\
\hline
\hline
1 & -0.15  & 0.042 & 0.18 & 0.60 & 0.044  & 0.61 & -2.33\\
1+2 & -0.15 & 0.043 & 0.23 & 0.60 & 0.042  & 0.42& -2.88\\
1+2+3 & -0.3 & 0.049 & 0.36 & 0.6 & 0.033  & 0.36& -3.54\\
1+2+3+4 & -0.4 & 0.052 & 0.47 & 0.4 & 0.031  & 0.49 & -3.55 \\
\end{tabular}
\captionof{table}{\label{m16_fit_param} Parameters of the fits of
the PDFs in Fig. \ref{M16_pdf}.}
\centering
\includegraphics[trim=0 0cm 2cm 0,width=0.48\linewidth]{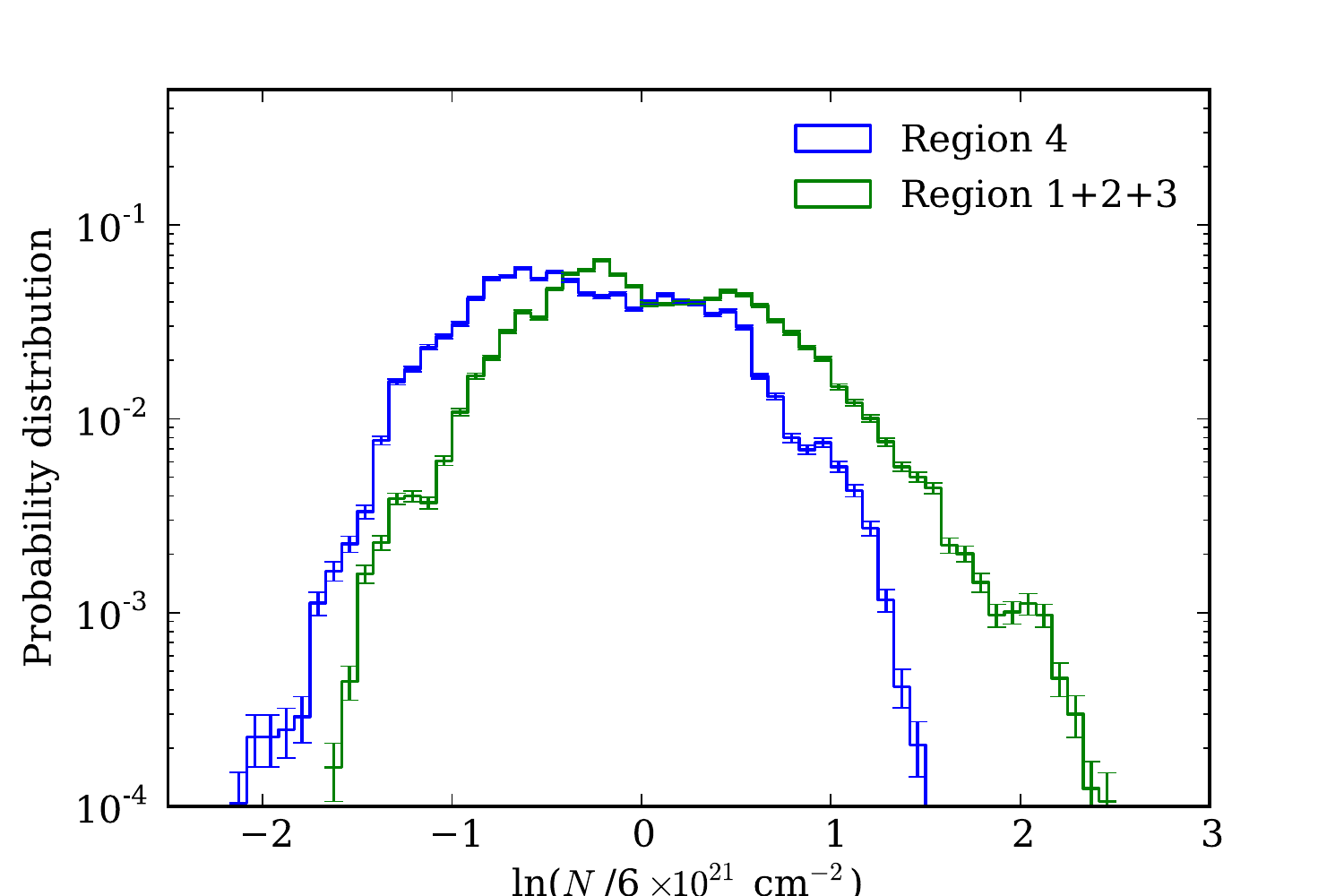}
\includegraphics[trim=0 0cm 2cm 0,width=0.48\linewidth]{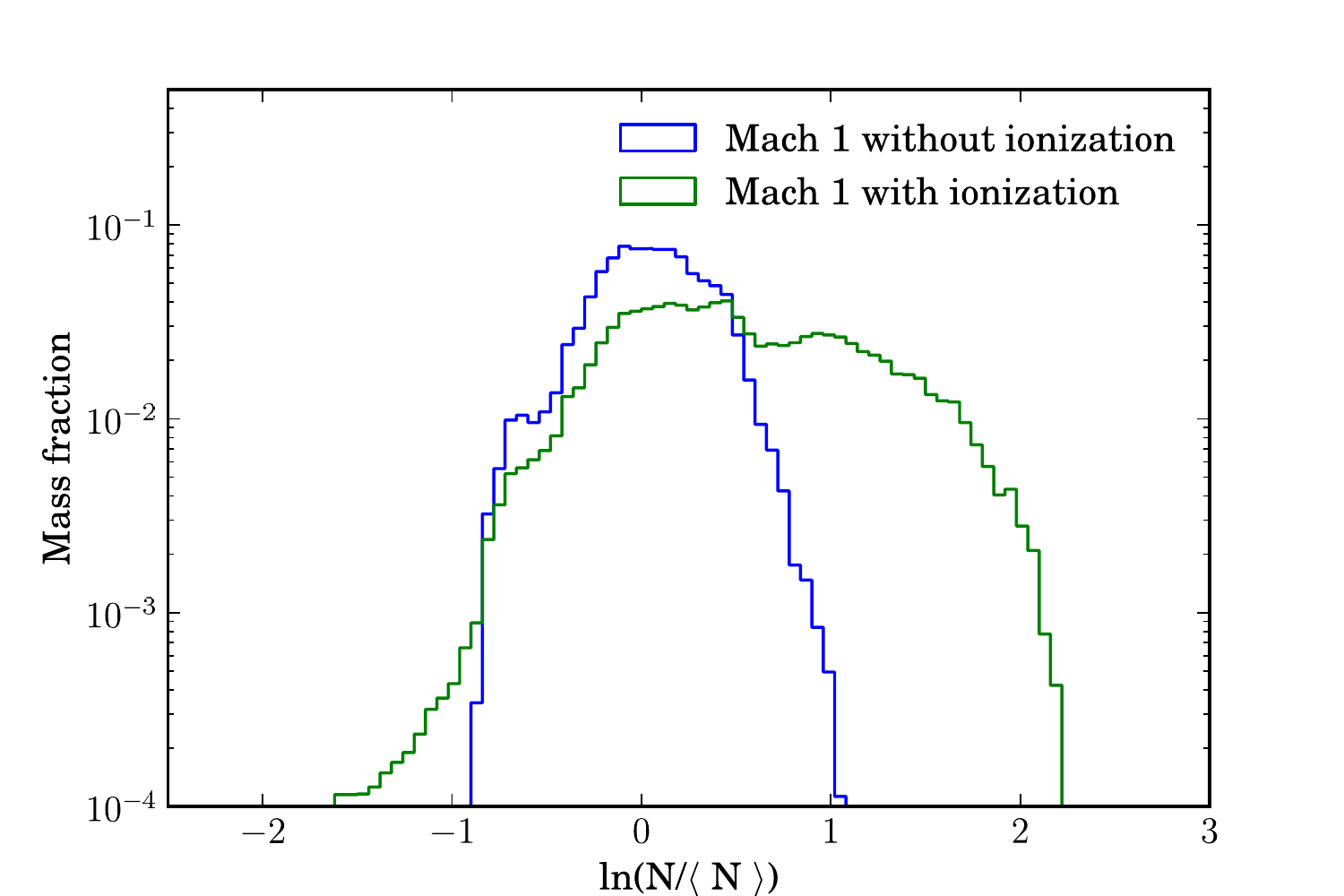}
\captionof{figure}{\label{M16_inout} Left: M16 column density PDFs on region 1+2+3 and region 4
  in Fig. \ref{M16_N_pdf}. Right: Column density PDFs of the simulation of
  the ionization of a Mach-1 turbulent cloud (without self-gravity). The excess at
  high-density and the double-peaked distribution can be identified
  both in observations and numerical simulations.}
\end{figure}

Figure \ref{M16_N_pdf} shows the different subregions in
  M16 used for the PDFs plotted in Fig. \ref{M16_pdf}. While the
  statistical Poisson error for the 
  distribution is low (error 
  bars are shown for the different PDFs) due to the large pixel
  statistics, there are systematic errors that influence the shape of
  the PDF.  First, the assumption of isothermality along the line of
  sight for fitting the
  pixel-to-pixel SED to derive the column density is not fully
  justified. Clouds with feedback due to radiation clearly show
  temperature variations. For M16, a temperature gradient from $\sim$23 K to
  $\sim$16 K was found \citep{Hill:2012jb}. 
  Second, the opacity increases towards higher column 
  densities \citep{Molina:2012iv} which leads to a narrower PDF and a
  steeper slope of the power-law tail for high densities.
 It is important to exclude -- or at least to quantify -- possible
  line-of-sight (LOS) confusion that may result in a superposition of
  several PDFs from different clouds. This effect was already proposed
  by \citet{Lombardi:2006fx} for the Pipe molecular cloud. It is
  necessary to use complementary spectroscopic data (preferentially 
  atomic hydrogen or low-J CO) to well define the velocity range of
  the cloud emission. Examples of how to estimate the contribution of
  foreground and background clouds in {\sl Herschel} column density
  maps are given in \citet{RiveraIngraham:2013wv} and
  \citet{Russeil2013}. In the case of M16, we do not expect the maps
  to be strongly affected by LOS confusion based on the CO line
  emission obtained by \citet{White:1999ue}.

 Each disk contains the smaller ones, therefore, Fig. 
\ref{M16_pdf} shows the evolution of the shape of the PDF while
increasing the radius of the disk around the ionizing
sources. The first disk (region 1) contains the ionizing sources and the
  dense column density structures up to the massive young stellar
  object at the west of the ionizing sources (blue circle). The second
disk (region 1+2) extends up to the base of the pillars of creation, the third disk
 (region 1+2+3) includes the whole dense structure at the north-west of the cavity
\citep[northern filament in][]{Hill:2012jb}, and the fourth disk
(region 1+2+3+4) reaches
the edge of the map.
In all the regions 
  a double peak is clearly identified, especially in region 1+2+3. In
  order to locate the position of the compressed gas, we plot the PDF
  of region 1+2+3 and the PDF of region 4 in
  Fig. \ref{M16_inout}. This region is an annulus that represents 
  the outward parts of the cloud that should not be too much affected
  by the ionization. However a dense component is still present and
  corresponds to the eastern filament identified in
  \citet{Hill:2012jb}. Apart from this filament, most of the compressed material in the second
  peak in the PDF in region 1+2+3 is not present in region 4,
  therefore the compression is likely to be linked with the ionized
  gas. To illustrate this point, we show in
  Fig. \ref{M16_inout} the PDF 
  obtained from the numerical simulation presented in
  \citet{Tremblin:2012he}. This simulation corresponds to the
  ionization of a turbulent cloud at Mach 1 (without self-gravity), in
  which the ionized-gas 
  pressure is much greater than the turbulent ram pressure. The
  compression is therefore efficient and a second peak appears in the
  PDF of the column density of the gas. This simulation is not performed in
  the exact conditions of M16, that are not exactly known because of
  projection effects, however the double-peak is a global property both
  identified in these observations and numerical simulations. 

  Because of this compressed peak, the PDFs are fitted using two
  lognormal distributions:
\begin{equation}\label{eq_fit}
p(\eta)=\frac{p_0}{\sqrt{2\pi\sigma_0^2}}\exp\left(\frac{-(\eta-\eta_0)^2}{2\sigma_0^2}\right)+\frac{p_1}{\sqrt{2\pi\sigma_1^2}}\exp\left(\frac{-(\eta-\eta_1)^2}{2\sigma_1^2}\right)
\end{equation}
with $\eta=\ln( N_{H_2}/\langle N_{H_2}\rangle)$. $ N_{H_2}$ is the column density and
$\langle N_{H_2}\rangle$ is the average of the column density in the regions 1+2+3+4
 $\langle N_{H_2}\rangle$=6$\times10^{21}$ cm$^{-2}$ for M16 (note that
this value is different from the averaged value on the whole
map). We used $\eta$ with a natural logarithm to ease comparisons with
previous works \citep[e.g.][]{Federrath:2013ip}. 
$\eta_i$ and $\sigma_i$ are the peak value and standard deviation
of each component, and $p_i$ is the integral of each component. The
first lognormal component can be linked to the initial turbulent
molecular cloud while the second lognormal component corresponds to a
compression of the turbulent cloud whose source can be either
ionization compression, colliding flows, or even winds, supernovas, etc. 
The
high column-density part of the PDFs is fitted with a power law of
power $m$ which corresponds to an equivalent spherical density profile
of power $\alpha$ \citep[see][]{Federrath:2012em,schneider2013}
\begin{eqnarray}\label{eq_pl}
p(\eta) &\propto& N_{H_2}^{m}\cr
\rho(r) &\propto& r^{-\alpha},\quad \mathrm{with} \quad \alpha = -2/m + 1
\end{eqnarray}
Assuming spherical symmetry is a crude approximation for complex and large
regions. However, if the pixels contributing to the power-law tail do
belong to a single condensation this approximation is relatively
accurate. Therefore we give the $m$ values in the tables of fitted parameters
and only use the equivalent $\alpha$ values to discuss the results on
single condensations or compare with previous works that used this
conversion. A spherical self-gravitating cloud has an $\alpha$ value
between 1.5-2 while higher values indicate a 
steeper profile influenced by compression.   
We point out that, for some PDFs, it could be argued that another
choice of function (e.g. a lognormal and a two-step power law) could
also give a good 
fit to the data, especially in region 1 in the case of M16. We will 
discuss in further detail in Sect. 6 the differences and motivations for a
two-lognormal plus one power-law fit or a single lognormal plus a
broken two-step power-law fit.

\begin{figure}[t]
\centering
\includegraphics[trim=0 3cm 2cm 0,width=\linewidth]{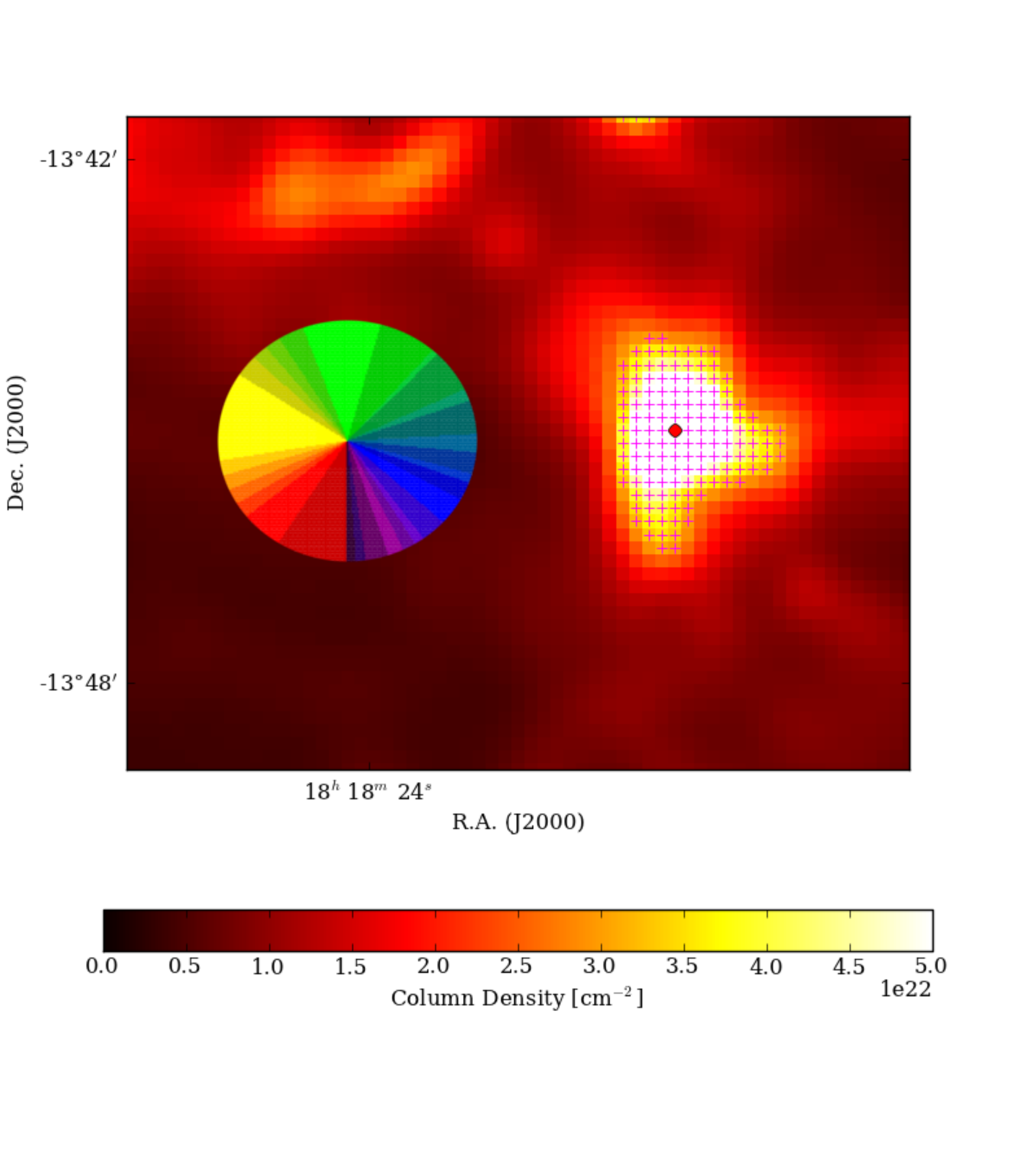}
\includegraphics[trim=0 0cm 0cm 0,width=\linewidth]{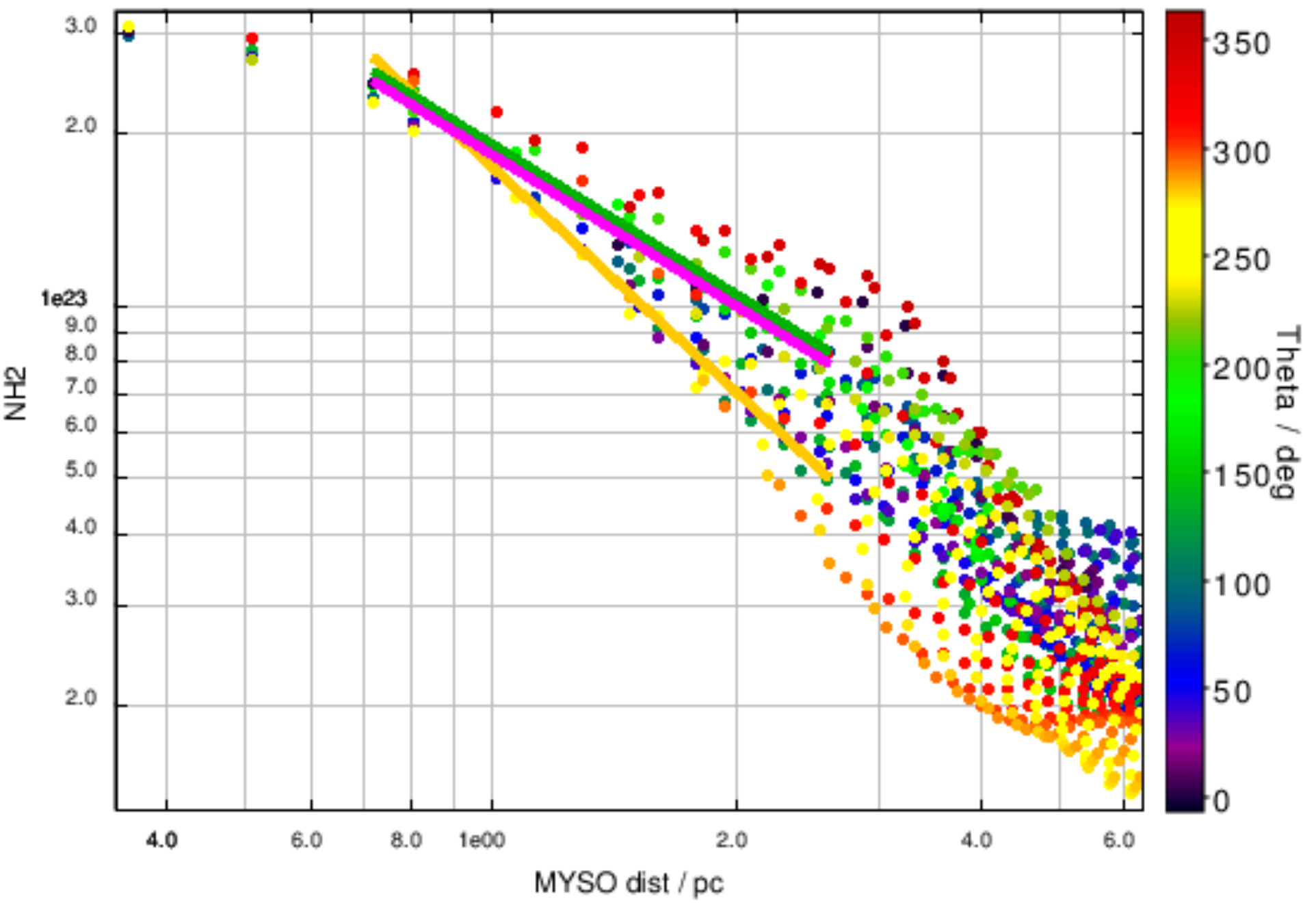}
\caption{\label{m16_myso} Top: zoom of the column density map on the
  western condensation. The magenta crosses indicate the pixels
  contributing to the power law of the PDF. The red dot is the
  position of the peak of the column density. The colormap is the
  color code used to show the orientation of the different
  profiles (e.g. green is north, yellow is east...). Bottom: Radial
  profiles of the column density of the 
  western condensation. The color indicates the orientation of the
  profile. The power-law fits are performed around three different
  directions: north (180$^\circ$)
  in green, east (270$^\circ$) in yellow and south (360$^\circ$) in purple. Each fit
  includes all the points in $\pm$ 40$^\circ$ around the main direction.}
\end{figure}

 Because of the number of parameters in Eq. \ref{eq_fit}, we fixed the
 peak values $\eta_0$ and $\eta_1$ based on the peak value of
 the low-density component seen in region 4 (Fig. \ref{M16_inout}) and
 the peak value of the compressed component seen in the concentric
 regions. After we have fixed these values $p_i$ and $\sigma_i$ are
 fitted to the observed distributions using a least-square method
 (Levenberg–Marquardt algorithm). We point out that the exact 
 values of the fitted parameters may depend on the choice of $\eta_0$ and
 $\eta_1$, however the general behavior deduced in our analysis is
 robust to this choice. The different parameters deduced from the fits are given in
  Table~\ref{m16_fit_param}. In M16, the
lognormal fit of the low column density (red-dashed line in
Fig. \ref{M16_pdf}) cannot account for the high column densities, and
the double peak is clearly visible even at high radius (e.g. region
1+2+3). The integral of the low-density peak $p_0$
increases with radius and the one of the compressed peak $p_1$ decreases
with distance. This behavior suggests that the compression effect is
more important close to the ionizing source and is indeed linked to the
ionization. Furthermore, the decrease in the compressed-peak gas quantity
is rather small even for 
region 1+2+3+4: the impact of the \ion{H}{ii} region is a
large-scale compression.

\begin{figure}[t]
\centering
\includegraphics[trim=0 3cm 2cm 0,width=\linewidth]{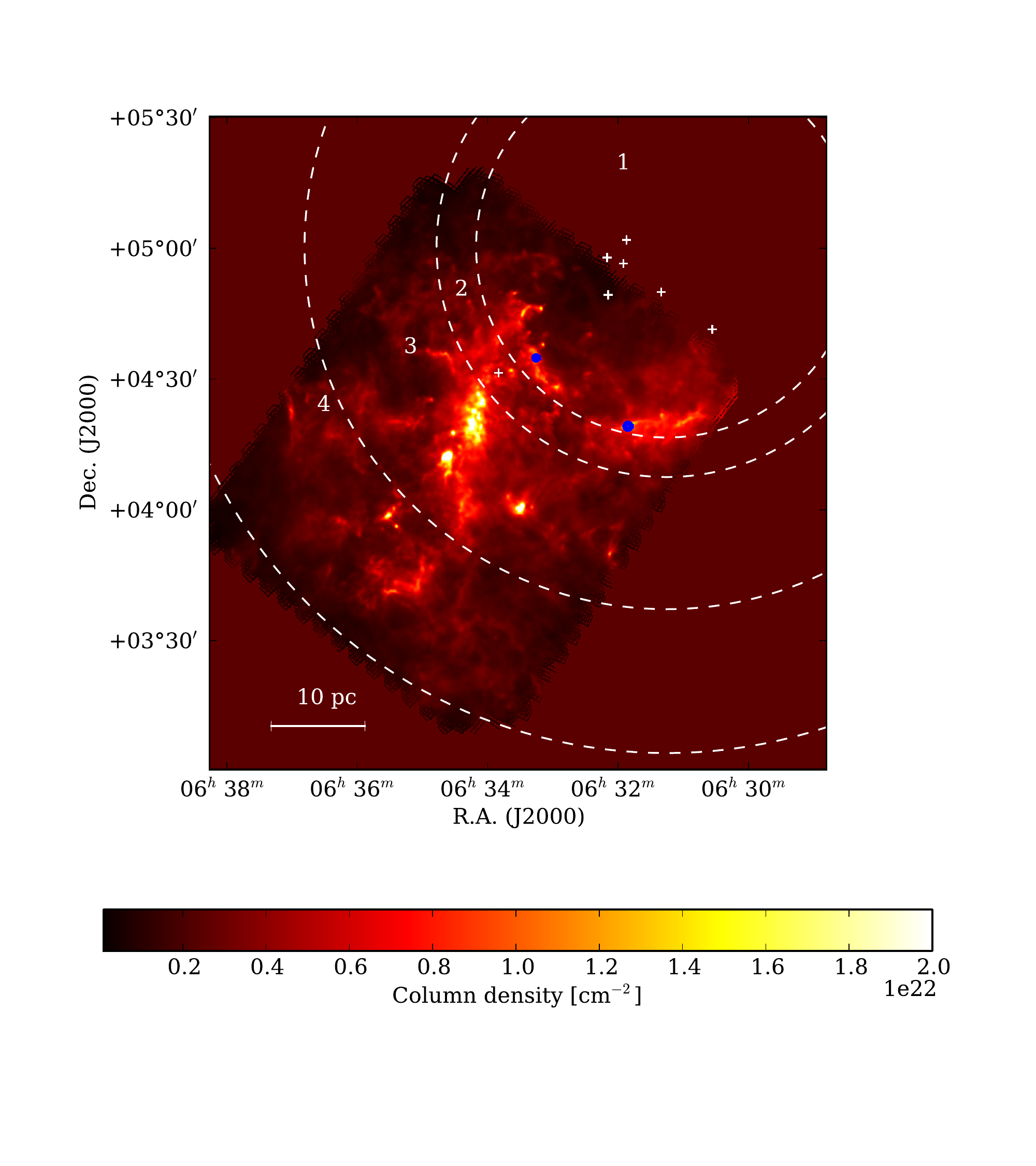}
\caption{\label{rosette_N_pdf} Rosette {\it Herschel} column density
  map \citep{Schneider:2012ds}. The circles indicate the different
  regions used for the PDFs. The white crosses indicate the position of
  the main ionizing sources, i.e. the most massive OB-stars from the
  NGC2244 cluster. The extension of the H$\alpha$ emission corresponds
to region 1+2 but there is also a significant FUV radiation up to
region 1+2+3 (see Schneider et al. in prep.). The blue circles
  indicates the position of the condensations 
  identified in the PDF of region 1 and region 1+2.}
\end{figure}

\begin{figure}[t]
\centering
\includegraphics[trim=0 0 1cm 0,width=\linewidth]{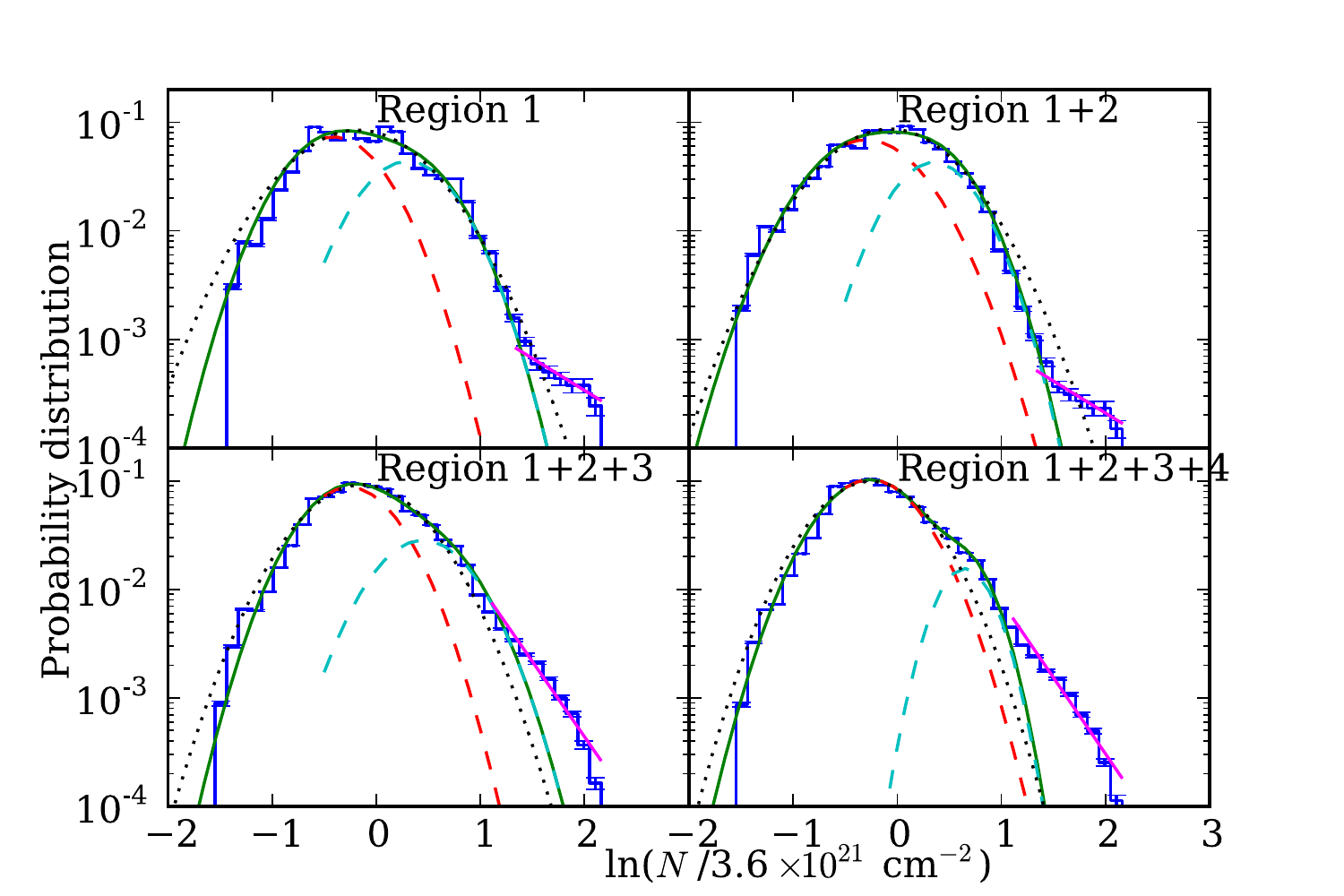}
\captionof{figure}{\label{rosette_pdf}  Rosette PDFs of the four regions
  indicated in Fig. \ref{rosette_N_pdf}. The multi-component fit is done using
  lognormal distributions (see Eq. \ref{eq_fit}) and a power law at
  high column densities (see Eq. \ref{eq_pl}). The black dotted line
  corresponds to a one-lognormal fit.}
\textcolor{white}{.}\\
\centering
\begin{tabular}{l|ccccccccc}
Region & $\eta_0$ & $p_0$ & $\sigma_0$ & $\eta_1$ & $p_1$ & $\sigma_1$ 
& $\chi^2_{2L}$ \\
\hline
\hline
1       & -0.41 & 0.07 & 0.39 & 0.30 & 0.04 & 0.38 & 250 \\
1+2     & -0.30 & 0.08 & 0.45 & 0.35 & 0.04 & 0.35 & 173 \\
1+2+3   & -0.26 & 0.09 & 0.39 & 0.45 & 0.03 & 0.40 & 236 \\
1+2+3+4 & -0.26 & 0.10 & 0.41 & 0.65 & 0.01 & 0.24 & 474 \\
\end{tabular}
\captionof{table}{\label{rosette_fit_param1} Parameters of the
  two-lognormal fits of
  the PDFs in Fig. \ref{rosette_pdf} and the corresponding reduced
  $\chi^2$ values.}
\textcolor{white}{.}\\
\centering
\begin{tabular}{l|cccccc}
Region & $\eta$ & $p$ & $\sigma$ & $\chi^2_{1L}$ & $m$\\
\hline
\hline
1       & -0.17 & 0.12 & 0.55 & 329  & -1.36 \\
1+2     & -0.07 & 0.12 & 0.53 & 210  & -1.36 \\
1+2+3   & -0.13 & 0.11 & 0.49 & 430  & -3.19 \\
1+2+3+4 & -0.25 & 0.11 & 0.45 & 1411 & -3.23 \\
\end{tabular}
\captionof{table}{\label{rosette_fit_param2} Parameters of the
  one-lognormal fits of
  the PDFs in Fig. \ref{rosette_pdf} with the corresponding reduced $\chi^2$
  values, and the exponents of the power-law fits at high column
  densities.}
\centering
\includegraphics[trim=0 0cm 2cm 0,width=0.48\linewidth]{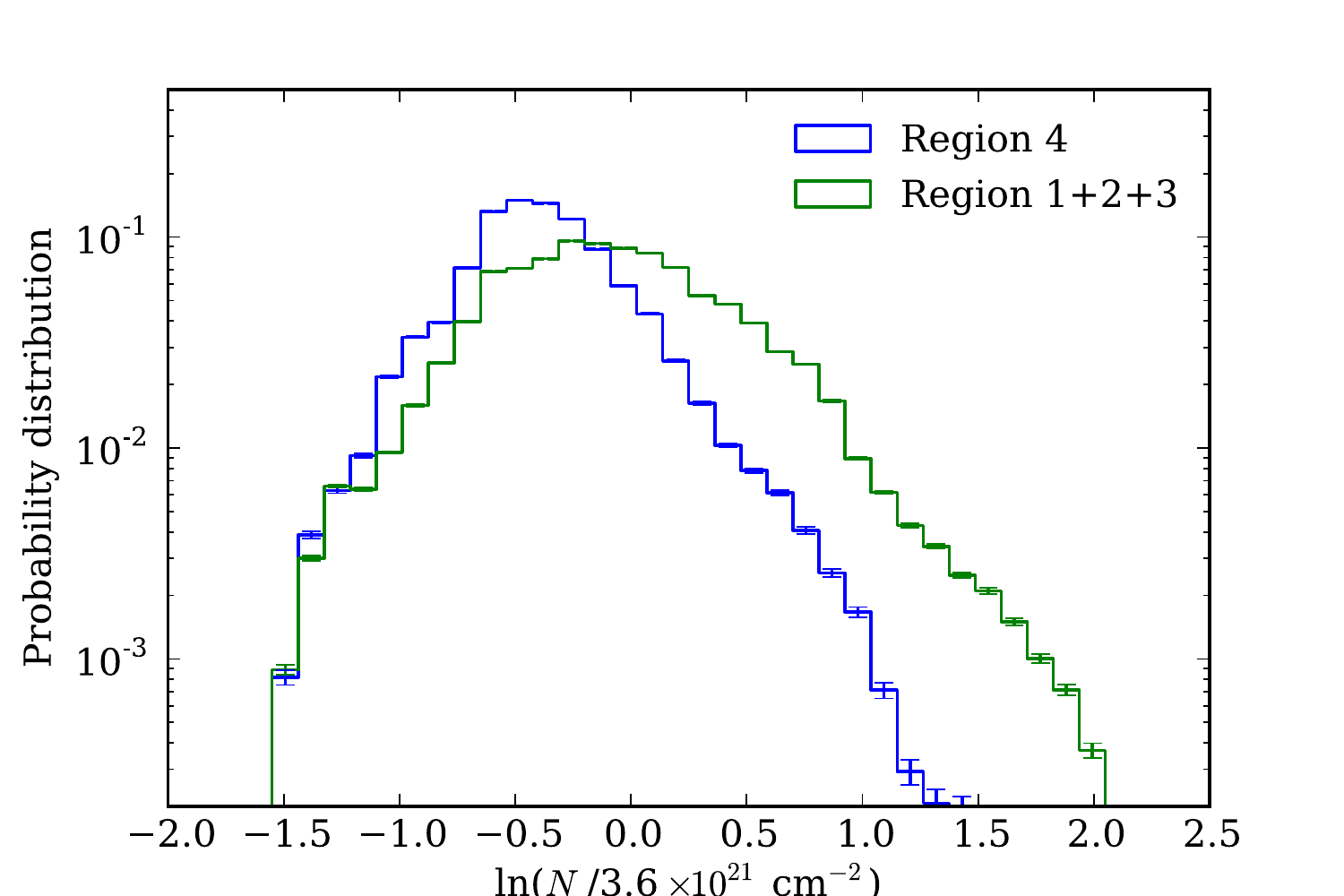}
\includegraphics[trim=0 0cm 2cm 0,width=0.48\linewidth]{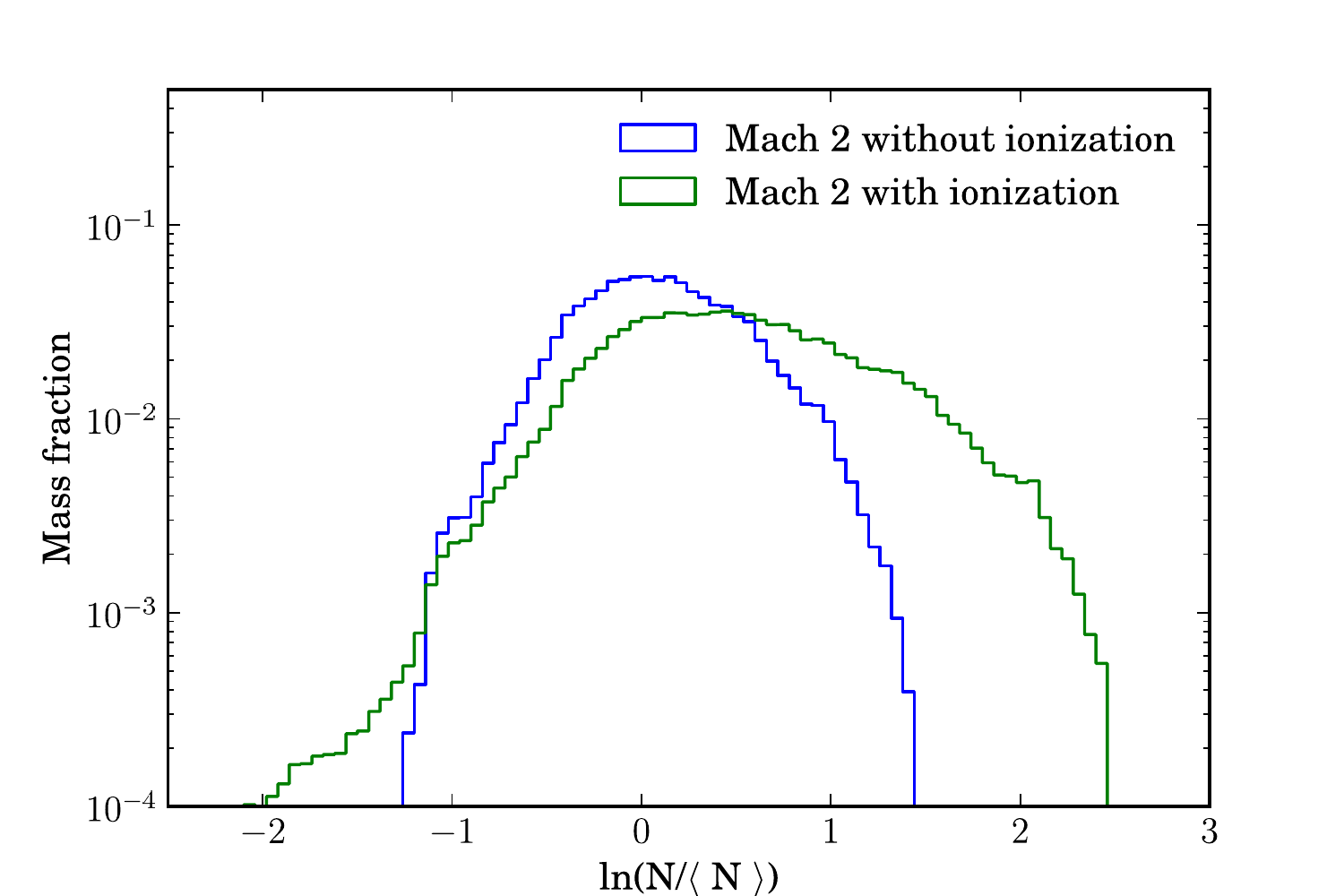}
\captionof{figure}{\label{rosette_inout} Left: Rosette column density PDFs on region 1+2+3 and region 4
  in Fig. \ref{rosette_N_pdf}. Right: Column density PDFs of the simulation of
  the ionization of a Mach-2 turbulent cloud. The excess at
  high density is still visible however the two peaks merge to form an
  enlarged distribution. The simulations are without self-gravity,
  therefore there are no power-law tails in the simulated PDFs while
  they are present in the observations. }
\end{figure}

In principle, a second compressed lognormal peak can be linked to shock
  compression around an \ion{H}{ii} region. This
  shock is likely to be driven by the expansion of the ionized gas.
  Gravitational instabilities -- 
  assuming that these are the most important physical processes for
  shaping the PDF -- lead to a power-law tail and could not account
  for the compressed peak visible in region 1+2+3. The power-law tail
  can still be identified at higher column densities (starting approximately for
  $\ln( N_{H_2}/\langle N_{H_2}\rangle) > 1.5$) where the distribution starts to
  deviate from the second log-normal. The slope fit leads to
  $\alpha$=1.86$\pm$0.05 in region 1 and decreases to
  $\alpha$=1.56$\pm$0.02 in region 1+2+3+4. These values are similar
  to those obtained in other regions like Aquila and Polaris
  \citep{schneider2013}. 
  In region 1, the pixels that contribute to the power law are part of
  two dense clumps indicated by blue circles in
  Fig. \ref{M16_N_pdf}. Although the radial profiles of these clumps
  do show signs of compression, this effect is not reflected in
  the PDF, which mixes both condensations in the power-law
  tail. We give as an example in Fig.~\ref{m16_myso} the radial profiles of the massive
  western young stellar object (MYSO hereafter) as a function of the
  orientation.
  The profile is very steep on the side facing ionization (east side
  in yellow) with an $\alpha$ value of 2.3 (2.56 when the background
  column density is subtracted), while the profiles in the
  south and north directions (red and green) have an $\alpha$ value of
  1.88 (1.99 when the background column density is subtracted) consistent with the
  influence of gravity. The average value is 2.25$\pm$0.02 suggesting that the
  compression on the MYSO is dominant. The same applies to the second condensation
  although the average $\alpha$ value is $\alpha$=1.76$\pm$0.02
  suggesting that gravity is dominant. 
Thus, $\alpha$
  values derived from the PDFs have to be taken with caution when different spherical
  structures are mixed in the same PDF. However when the PDF is
  computed on separated areas for the 
  two condensations, the deduced $\alpha$ values are compatible with
  their averaged fitted profiles. The use of steeper profiles to identified
compressed cores will be discussed in Sect.~6.

\section{Rosette Nebula}\label{sect_rosette}

 The Rosette molecular cloud is located at 1.6 kpc from the Sun
  \citep{Williams:1994hla,Schneider:1998ta,Heyer:2006hv} and
  is associated with a prominent \ion{H}{ii} region, illuminated by
  the central cluster NGC 2244.  The 17 OB stars of the cluster have a
  total Lyman-$\alpha$ luminosity of 3.8$\times$10$^5$ L$_\odot$
  \citep[see][]{Cox:1990ub}. The UV-flux is of the order of 100-200
  G$_\circ$ (Habing-field) \citep{Schneider:1998vx} at the interface
  zone between \ion{H}{ii} region and molecular cloud and decreases to
  a few G$_\circ$ for the bulk emission deep ($\sim$30 pc) into the
  cloud.  Photon dominated regions (PDRs) are found at the cloud
  surfaces where the UV-radiation shines directly onto the molecular cloud
  \citep[see][]{Schneider:1998ta,Schneider:1998vx}.  It is there where
  many pillars and globules are formed \citep[see][and Tremblin et
    al. in press]{Schneider:2010ec}. Although the total UV-field is low
  (on average only $\sim$10 G$_\circ$, see Table~\ref{hobys}) --
  compared to high-UV regions like Carina or Cygnus with up to 10$^5$
  G$_\circ$ -- the very clumpy and filamentary structure of the cloud
  enables a large penetration depth of UV-radiation so that PDRs are
  also found deep inside the cloud. The Rosette cloud contains several
  embedded infrared-clusters \citep[see][and references
  therein]{Poulton:2008fp}, and is actively forming stars, indicated by
  a larger reservoir of clumps \citep{DiFrancesco:2010ef}, a large
  population of protostars \citep{Hennemann:2010hp}, molecular
  outflows (White et al. in prep.), and a few
  candidate massive dense cores \citep{Motte:2010fy}. There is a
  long ongoing-discussion whether star formation is {\sl triggered} in
  Rosette, with a possible increasing age gradient from the ionization front into
  the cloud \citep[see][for a detailed discussion]{Schneider:2010ec}. For Rosette, the bulk emission of the cloud is
  at $\approx$16$\pm$6 km s$^{-1}$ 
  \citep{Williams:1994hla,Schneider:1998ta} and there is no
  significant emission neither in HI nor CO 
  outside this velocity range. Thus, the column density map is
  probably not very much affected by LOS confusion.

\citet{Schneider:2012ds} has already shown that a double peak in the PDF
was found for the gas close to the border of the shell ($\approx$ 20
pc of radius). However, their distinctions into subregions were
  mainly based on morphology, so that here, we study the evolution of
  the PDF as a function of radius around 
the \ion{H}{ii} region in a more
systematic way, by taking the distribution in concentric
  disks around the bubble. In Fig. \ref{rosette_N_pdf}, we show the
four different regions that are considered for the PDFs presented in
Fig. \ref{rosette_pdf}. The first disk (region 1) contains the ionizing sources and the
  dense column density structures up to the massive young stellar
  object at the south of the ionizing sources (blue circle). The second
disk (region 1+2) extends up to the northern part of the main star-forming region
of Rosette \citep[see][]{Schneider:2012ds}, the third disk (region 1+2+3)
includes this star-forming region, and the fourth disk (region
1+2+3+4) reaches
the edge of the map. 
In Fig. \ref{rosette_inout}, we show the PDF
of the region 1+2+3 and the annular region 4 in order to compare the
compressed part that is influenced by an effect localized near the
ionizing sources. The PDF of region 1+2+3 is not double-peaked as was seen for M16
and even the one of region 1 alone shows only a slight indication of a
double-peak, less clearly than was presented in
\citet{Schneider:2012ds}. Indeed, in \citet{Schneider:2012ds}, the
selection of 
regions focused explicitly on the interaction zone (only a small part
of our region 1) and thus revealed the double-peak. Taking a larger
area adds more pixels that are not directly impacted by the ionization
front (towards the south-west in our region 1) and thus dilutes the double-peak.
For comparison we show also in
Fig. \ref{rosette_inout} the PDFs of a simulated Mach-2 turbulent
cloud exposed to ionization \citep{Tremblin:2012he}. The PDF presents
a similar shape where 
the two peaks merge to form an enlarged distribution. One could argue
that these enlarged distributions can be fitted properly by a single
lognormal component. We show in Fig. \ref{rosette_pdf} such a fit
(black-dotted line) and the corresponding parameters are given in
Table~\ref{rosette_fit_param2} while the parameters of the
two-lognormal fit are given in Table~\ref{rosette_fit_param1}. The
reduced $\chi^2$ values for both fits are computed assuming a Poisson
noise in each bin. The corresponding error-bars are plotted in
Fig. \ref{rosette_pdf} but the pixel statistic is so large that the
error-bars are small and barely visible in log-space. Such small
error-bars explain the large values of the $\chi^2$ for both
fits. This means that these regions have complex structures and
physics that are not well represented by the fitting models. We
performed a F-test to check that the two-lognormal fit provides a
significant improvement \citep[see][]{Bevington:2003tc}. The F values
for the different regions are 
respectively 3.3, 2.6, 6.8, and 15 and the critical value is 3 for a
false-rejection probability of 5 \%. Therefore, 
the two-lognormal model gives a significantly better fit to the data especially in
regions 1+2+3 and 1+2+3+4.
 In all regions, there is an excess of compressed gas compared to
 the PDF in the annular region 4. 
   The integral of the compressed component $p_1$ in
   Table~\ref{rosette_fit_param1} 
decreases from 0.04 to 0.01 when the radius of the region increases.
This behavior is expected: when the radius of the region increases,
more and more unperturbed gas is added in the distribution while the
compressed gas remains the same. Therefore the relative importance of
the compressed gas decreases.

When this excess is
   taken into account by a second lognormal component in the fits in
   Fig. \ref{rosette_pdf}, a power-law deviation from the two-lognormal
   distribution is still visible and indicates the possible influence of
   gravity but at higher column densities. It is
   especially clear in regions 1 and 1+2 that the power-law deviation
   is occurring around $\eta$ $\approx$ 1.5, indicating the regime in which
   gravity starts to dominate. Assuming an equivalent spherical density distribution
  ($\rho \propto r^{-\alpha}$), 
   a transition from an exponent $\alpha$ of 2.38$\pm$0.10 in regions 1
   and 1+2 to a
   value of 1.62$\pm$0.08 in region 1+2+3+4 is derived from the
   power-law fits. This transition indicates that the ionization can have
   an influence even on the gravity-dominated regions that are in the
   PDR. Assuming spherical 
  symmetry on large scale (regions 1+2+3 and 1+2+3+4) is a very crude approximation in these
  complex regions and a detailed analysis taking into account the Mach
  numbers is needed to study the interplay between other different processes.  
  In region 1 and 1+2, the pixels that have a column
  density with $\eta$ higher 
  than 1.5 correspond to two nearly spherical dense condensations in
  the Monoceros Ridge 
  and in south dense region at the edge of the cavity (blue circles
  in Fig. \ref{rosette_N_pdf}) 
  which were already identified as regions of gas compression at the
  cloud-nebula interface \citep{RomanZuniga:2008tm}. As in M16, these
  $\alpha$ values have to be taken with caution since they mix
  different structures. Nevertheless, the high alpha
  value observed in regions close to the ionizing sources was already
  seen in other regions like the central, dense, high-mass
  star-forming ridge in the cloud NGC 6334 \citep{Russeil2013} and
  more recently in \citet{Palau:2013} (for the massive dense cores
  DR21-OH, AFGL 5142 and CB3). The 
  transition from high values to low values typical of other clouds
  seems to indicate that the compression from ionization is also
  playing a role at high column densities where structures are
  expected to be gravitationally collapsing.

\section{RCW 120}\label{sect_rcw120}

\begin{figure}[t]
\centering \includegraphics[trim=0 3cm 2cm 0, width=\linewidth]{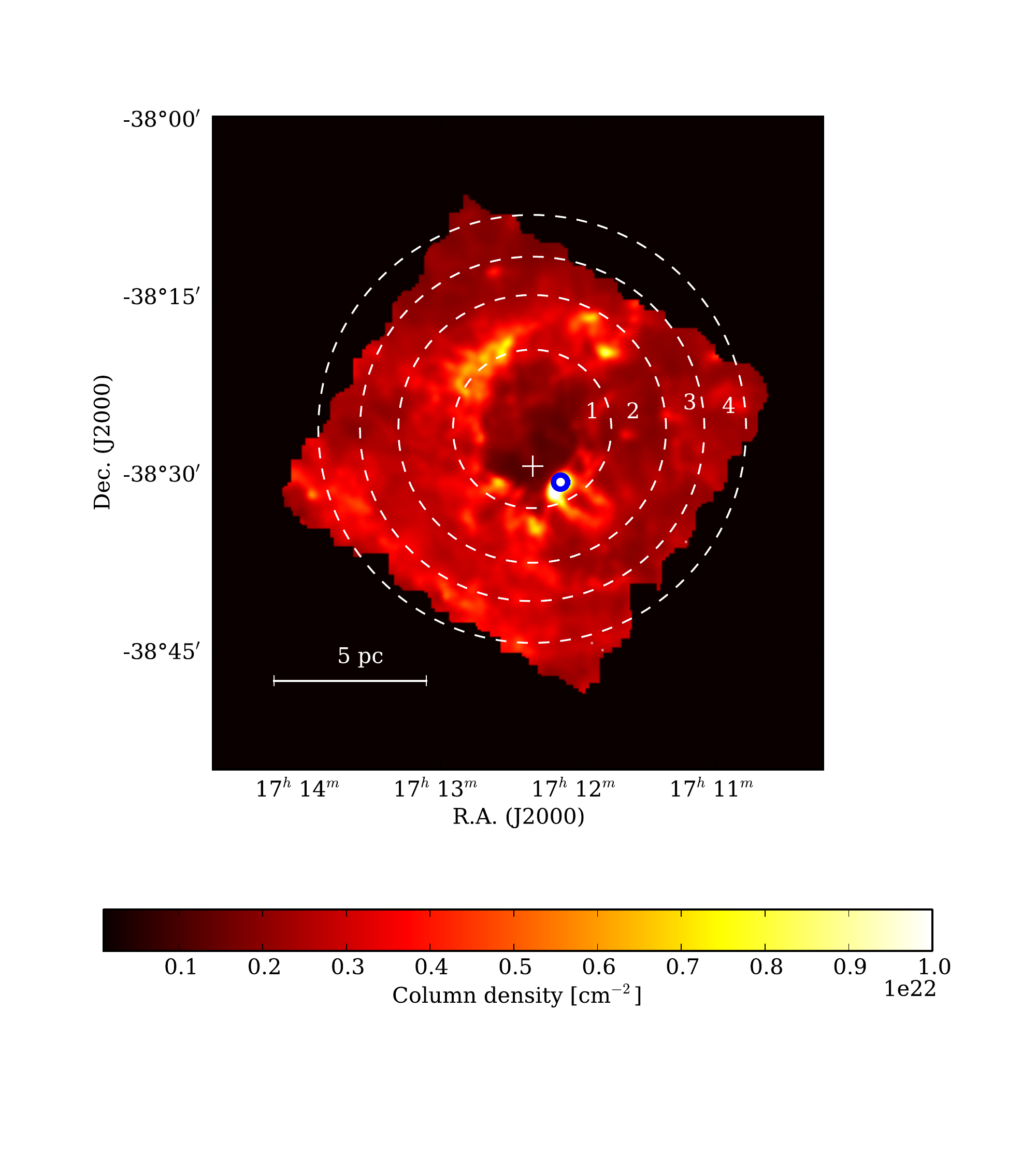}
\caption{\label{rcw120_N} RCW 120 {\it Herschel} column density map
  \citep{Anderson:2012jy}. The circles indicate the different regions
  used for the PDFs. The white cross indicates the position of the
  main ionizing source (O8 star). The extension of the H$\alpha$
  emission and the limit of the \ion{H}{ii} region is inside region
  1. The blue circle
  indicates the position of the possible compressed core collapse
  identified in the PDFs.}
\end{figure}

 RCW 120 (Fig.~\ref{rcw120_N}) is an \ion{H}{ii} region
\citep[see][]{Rodgers:1960tc,Zavagno:2007ix} located at a distance of 1.3 kpc from
the Sun and well studied by {\it Spitzer} \citep{Deharveng:2009kd} and
{\it Herschel} \citep{Zavagno:2010jv,Anderson:2010dp}. Its
circular shape led to the name of ``the perfect bubble''
\citep{Deharveng:2009kd}, although the ionizing source, an O8 star
\citep{Martins:2010jw}, is not at the centre of the ionized sphere.
Since the ionizing source is very close to the southern part of the
bubble, it is probable that the expansion is asymmetric. The molecular
gas is denser at the south than the north of the bubble, as indicated
by the column density (see Fig.  \ref{rcw120_N}). Because of the asymmetry, the 1D
model of collect and collapse is difficult to apply in the form
presented in \citet{Elmegreen:1977iq}. Among other regions, the dust
properties of RCW 120 has been 
recently analyzed by \citet{Anderson:2012jy} using {\it Herschel}
observations. These authors found that the mass of the material in the
PDR is compatible with the expected mass swept up during the expansion
of the \ion{H}{ii} region and that the FIR emission suggests that the
bubble is a three-dimensional structure. We used in the present
analysis the column density map derived by \citet{Anderson:2012jy}. We
point out that the column density map of 
  RCW120 is done by a fit in linear space rather than log space (as done for the
  other {\it Herschel} maps). However we do not expect this
  inconsistency to influence our results.

\begin{figure}[t]
\centering \includegraphics[trim=0 0 1cm 0, width=\linewidth]{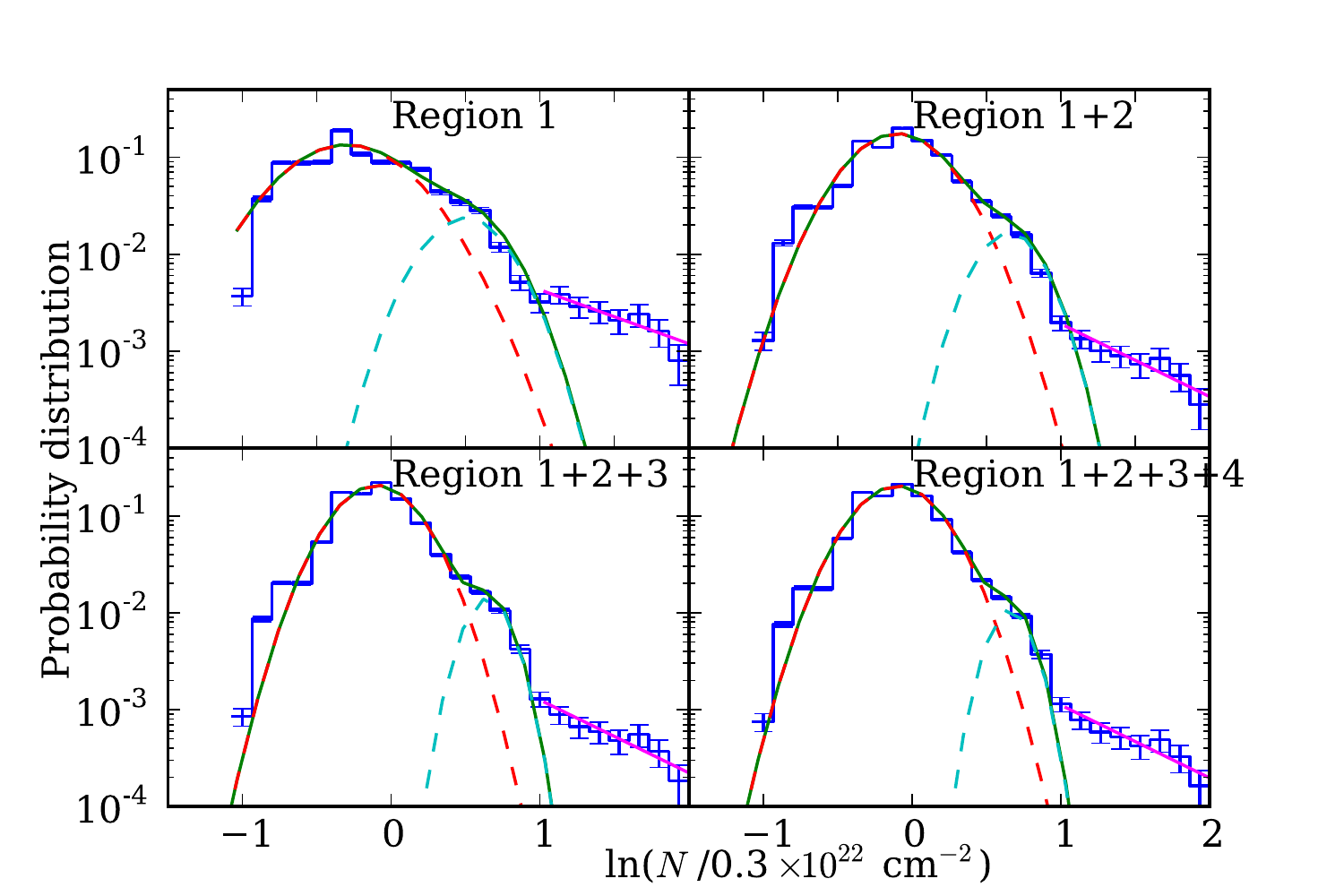}
\captionof{figure}{\label{rcw120_pdf} RCW 120 PDFs on the four regions indicated
  in Fig. \ref{rcw120_N}. The multi-component fit is done using
  lognormal distributions (see Eq. \ref{eq_fit}) and a power law at
  high column densities (see Eq. \ref{eq_pl}).}
\textcolor{white}{.}\\
\begin{tabular}{l|ccccccc}
Region & $\eta_0$ & $p_0$ & $\sigma_0$ & $\eta_1$ & $p_1$ & $\sigma_1$ & $m$
\\ \hline \hline 
1       & -0.3 & 0.12 & 0.36 & 0.5  & 0.014 & 0.24 & -1.28\\ 
1+2     & -0.1 & 0.12 & 0.29 & 0.65 & 0.008 & 0.19 & -1.73\\ 
1+2+3   & -0.1 & 0.13 & 0.25 & 0.65 & 0.005 & 0.14 & -1.73\\ 
1+2+3+4 & -0.1 & 0.13 & 0.13 & 0.65 & 0.003 & 0.13 & -1.73\\
\end{tabular}
\captionof{table}{\label{rcw120_fit_param} Parameters of the fits of
  the PDFs in Fig. \ref{rcw120_pdf}.}
\end{figure}

\begin{figure}[t]
\centering
\includegraphics[trim=0 3cm 2cm 0,width=\linewidth]{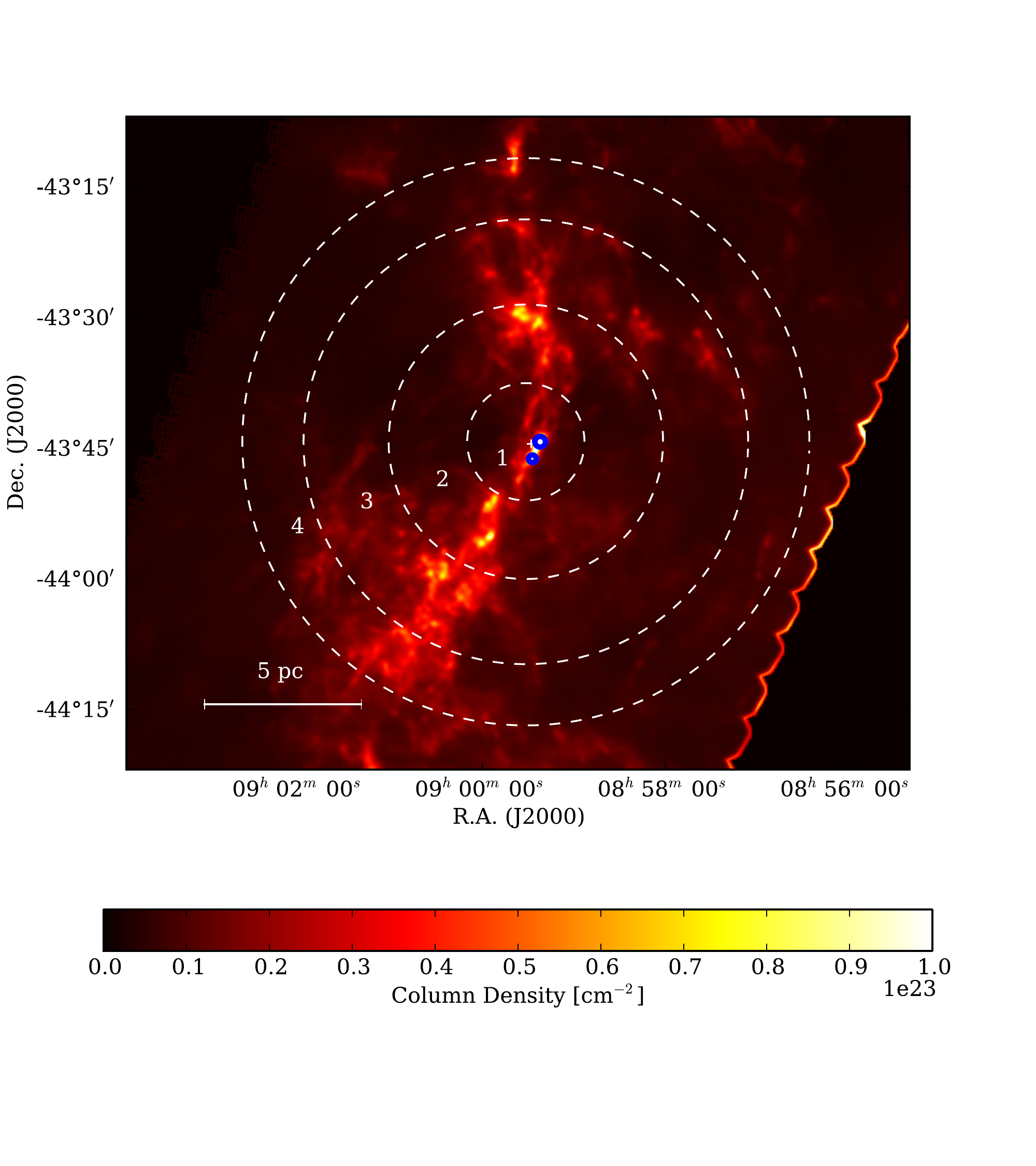}
\caption{\label{vela_N_pdf} RCW 36 {\it Herschel} column density
  map \citep{Minier:2013ih}. The circles indicate the different regions used for the
  PDFs. The white cross indicates the position of the main ionizing
  source. The limit of the H$\alpha$
  emission and the limit of the \ion{H}{ii} region are inside region
  1. The blue circles
  indicate the positions of the condensations
  identified in the PDF of region 1.}
\end{figure}
\begin{figure}[t]
\centering
\includegraphics[trim=0 0cm 1cm 0,width=\linewidth]{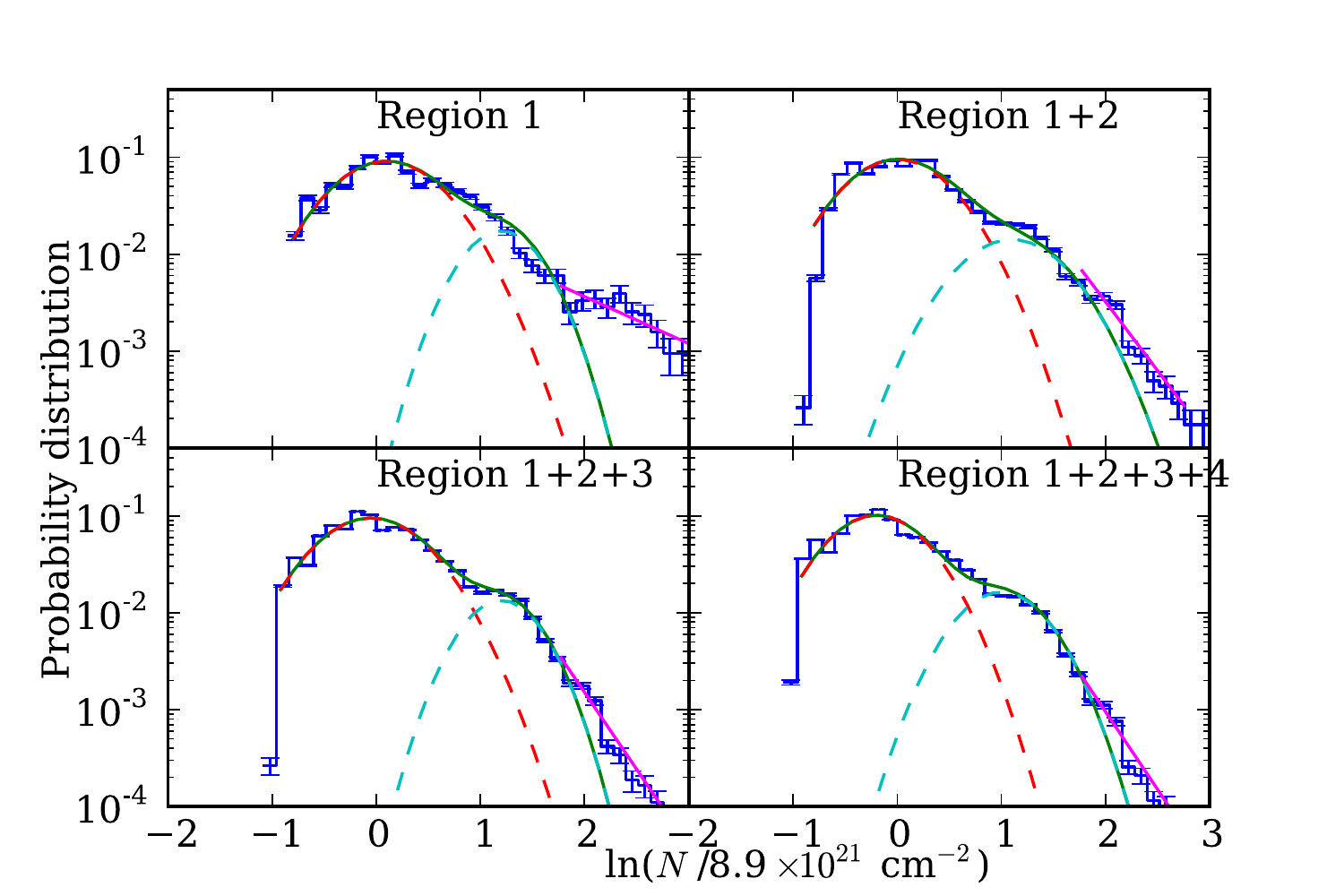}
\captionof{figure}{\label{vela_pdf} RCW 36 PDFs on the four regions indicated
  in Fig. \ref{vela_N_pdf}. The multi-component fit is done using
  lognormal distributions (see Eq. \ref{eq_fit}) and a power law at
  high column densities (see Eq. \ref{eq_pl}).}
\textcolor{white}{.}\\
\centering
\begin{tabular}{l|ccccccc}
Region & $\eta_0$ & $p_0$ & $\sigma_0$ & $\eta_1$ & $p_1$ & $\sigma_1$ & $m$\\
\hline
\hline
1       & 0.1   & 0.11 & 0.47 & 1.2  & 0.014 & 0.33 & -1.12\\
1+2     & 0.    & 0.11 & 0.45 & 1.1  & 0.016 & 0.45 & -3.28\\
1+2+3   & -0.05 & 0.11 & 0.31 & 1.2  & 0.011 & 0.33 & -3.65\\
1+2+3+4 & -0.2  & 0.11 & 0.42 & 1.0  & 0.016 & 0.38 & -3.70\\
\end{tabular}
\captionof{table}{\label{vela_fit_param} Parameters of the fits of
the PDFs in Fig. \ref{vela_pdf}.}
\end{figure}

In Fig. \ref{rcw120_N}, we show the four different regions that are
considered for the PDFs presented in Fig. \ref{rcw120_pdf}. The
parameters of the fits are listed in Table~\ref{rcw120_fit_param}.
The PDFs are well fitted by a two-component
lognormal and the excess in the compressed component is clearly
visible in regions 1+2+3 and 1+2+3+4. At high column densities, it is
possible to identify the 
dense clump formed at the south west of the ionizing source in the
dense part of the PDF (blue circle in Fig. \ref{rcw120_N}). This clump is not part of the compressed
component and the PDF is well fitted by a power-law suggesting the
role of gravity. In region 1, the power-law fit leads to an exponent $\alpha$ of
2.56$\pm$0.17 that may indicate the role of ionization compression in
the formation of this condensation and its collapse
to form stars. The radial
profile of the condensation has an exponent $\alpha$
of 2.39$\pm$0.01, which is in a good agreement with the value deduced
from the PDF. 
If the gravitational collapse would
have happened before, we would expect the exponent to be close to 2
and to be only marginally affected by the ionization since the
ionizing front did not overwhelm the condensation. Indeed the
numerical study in \citet{Minier:2013ih} in the case of RCW 36 showed
that the collapse of a condensation that is located in the shell is likely to be
triggered. If the condensation had gravitationally collapsed prior to
the passage of the ionization front, the condensation will be already sufficiently
dense to resist the ionization front expansion. It would for example
trigger the formation of a pillar rather than a condensation remaining in the 
shell. Furthermore a   
statistical argument based on the study of \citet{Thompson:2012gn}
also suggests that the dense condensation is linked to the \ion{H}{ii}
region.  
Based on the numerical study of \citet{Minier:2013ih}, the triggering
can be confirmed by comparing the velocities of the
condensation and the nearby 
shell. If the velocities are the same,
the formation of the condensation is dynamically linked to the shell
and likely to be triggered.

The amplitude of the compressed lognormal decreases when the radius of
the region increases (see Table~6). This is consistent with the picture we got
from Rosette and M16: the larger the region, the
  less important becomes the peak, because more and more unperturbed
gas is added to the distribution. However, the decrease of the
amplitude of the compressed peak is more important in the case of RCW
120 (from 0.014 to 0.003).  This can be explained by the \ion{H}{ii}
region occupying a smaller part of the image than the large ionized bubbles in the Eagle and
Rosette Nebula.  The compression factor of the column density is
typically of the order of 2 (given by $\exp(\eta_1-\eta_0)$). That is rather
low compared to the factor 20-30 expected from the 1D simulations of
\citet{Zavagno:2007ix}. However the difference could be the result of
a projection effect. Considering the size of the
bubble which has a radius of 1.8 pc, we can assume that the
thickness along the line of sight of the compressed layer is of the order of 1 pc.
The cloud is at a column density of 10$^{23}$
cm$^{-2}$ and corresponds to a density of 3$\times$10$^{3}$ cm$^{-3}$
integrated over a line of sight of 10 pc. The compressed layer has a column density of around
$2\times10^{23}$ cm$^{-2}$ which corresponds to a density of
6$\times$10$^{4}$ cm$^{-3}$ integrated over 1 pc. Therefore we can get a
compression of $\approx$ 20 if the line-of-sight thickness of the
shell is ten times smaller than the thickness of the cloud. Therefore
a factor of compression of 20-30 in density can  
result in a factor of compression of 2 in column density.

\section{Vela C/RCW 36} \label{sect_rcw36}

RCW 36, at a distance of 700 pc from the Sun \citep{Murphy:1991ww},
 has the shape of an hourglass. At the centre, an embedded
cluster, 2-3 Myr old, has 350 members with the most massive star being
a type O8 or O9 \citep{Baba:2004gm}. The cluster extends over a radius
of 0.5 pc, with a stellar surface number density of 3000 stars
pc$^{-2}$ within the central 0.1 pc. FIR emission and radio continuum
emission are reported to be consistent with the presence of an
\ion{H}{ii} region \citep{Verma:1994vt}. The star cluster has probably
inflated the G265.151+1.454 \ion{H}{ii} region \citep{Caswell:1987tc}
that is responsible for the H$\alpha$ emission originally observed by
\citet{Rodgers:1960tc}. Looking at larger scales, RCW 36 is located
within the Vela C molecular cloud that consists of an apparent network
of column density filaments \citep{Hill:2011ht}. Such filaments are
star formation sites. Herschel observations of low-mass star-forming
molecular clouds \citep[e.g.][]{Arzoumanian:2011ho} show that stars
form in supercritical filaments while \citet{Schneider:2012ds} propose
that star-cluster formation sites correlate with filament
network junctions. Among the networks of filaments in Vela C, there is
a more prominent and elongated interstellar dust structure of
$\approx$ 10 pc in length that is named the Centre-Ridge by
\citet{Hill:2011ht}. It encompasses RCW 36, which is at its Southern
end, a bipolar nebula surrounded by a dense ring of gas in the plane
of the Centre-Ridge \citep{Minier:2013ih}.

 Figure \ref{vela_N_pdf} shows the {\it Herschel} column density
  map of RCW36 inside the Vela C molecular cloud with the different
  regions for which we determined the PDFs (Fig. \ref{vela_pdf} and
  Table~\ref{vela_fit_param}).  The first region 
corresponds to the extension of the ionized gas. The PDF presents two
components, however, contrary to the other regions, the amplitude of
the compressed peak does not decrease  with increasing radius. In region 1,
the high-density component consists of the dense ring close to the
cluster \citep{Minier:2013ih}, and the low-density one corresponds to
the low-density medium that is present on each side of the bipolar
nebula. The deviation at high densities ($\ln( N_{H_2}/\langle
N_{H_2}\rangle)\approx 2$) corresponds to the dense condensations that are
present in the ring and were also discussed in \citet{Minier:2013ih}.
  A power-law tail (although complex)
  for this range can be identified. Though the pixel statistic is
  small and the error larger, we
  tentatively determine a value of $\alpha$=2.80$\pm$0.23 which is
  much higher than 2. This high value is also consistent with the
  filamentary profile measured by \citet{Hill:2012ip}
  ($\alpha$=2.70$\pm$0.2). We checked the radial profiles of the densest and
  largest condensation and got an averaged fitted value of
  $\alpha$=2.30$\pm$0.11 (in some direction the profile has an alpha
  value of 2.83). 
We may have here
  the same situation as in RCW120 where the PDF also shows a clear
  excess for the highest densities due to {\sl individual compressed core
    collapse}. This supports the previous study in
  \citet{Minier:2013ih} showing that, based on the morphology of the
  regions, the gravitational collapse of these condensations (blue
  circles in Fig. \ref{vela_N_pdf}) was
  likely to be triggered (also confirmed by the recent age
  determination of the different stellar populations done by \citet{Ellerbroek:2013tv}). Furthermore, the exponent $\alpha$ decreases
  to a value of 1.54$\pm$0.02 in region 1+2+3+4. The region is so
  large that the dense small clumps contribute to the probability
  distribution at less than 10$^{-4}$ and therefore are not visible in
  the PDFs anymore. The dense compressed power-law tail is therefore only
  present in region 1 and 
  probably linked to the ionization.
 In previous large-scale studies, e.g. the
  Vela C cloud (Hill et al.  2011), Rosette (Schneider et al. 2012),
  and NGC6334 (Russeil et al. 2013), it was not possible to
  distinguish whether the power-law tail arises from
  collapse of many cores and/or global collapse of clumps or
  filaments/ridges since both processes play a role. In smaller
  regions, we are less perturbed by a large
  amount of bulk gas that constitutes the PDF and global collapse may
  play a minor role. However, we emphasize that this finding remains
  slightly speculative and needs further detailed studies.

 In any case, when the radius of the area increases, the
low-density component remains the same and the amplitude of the
high-density peak increases. This behavior is expected because
the influence of the \ion{H}{ii} region is limited to region 1 and in
the other regions, the
high-density component does not consist of the compressed ring 
anymore, but is dominated by the south-north elongated molecular cloud. When the
radius is high, the area of the compressed ring around the \ion{H}{ii}
region is very small compared to the rest of the molecular
cloud.

\section{Discussion and interpretation}

\subsection{From large-scale to small-scale ionization compression}

\begin{figure}[t]
\centering \includegraphics[width=\linewidth]{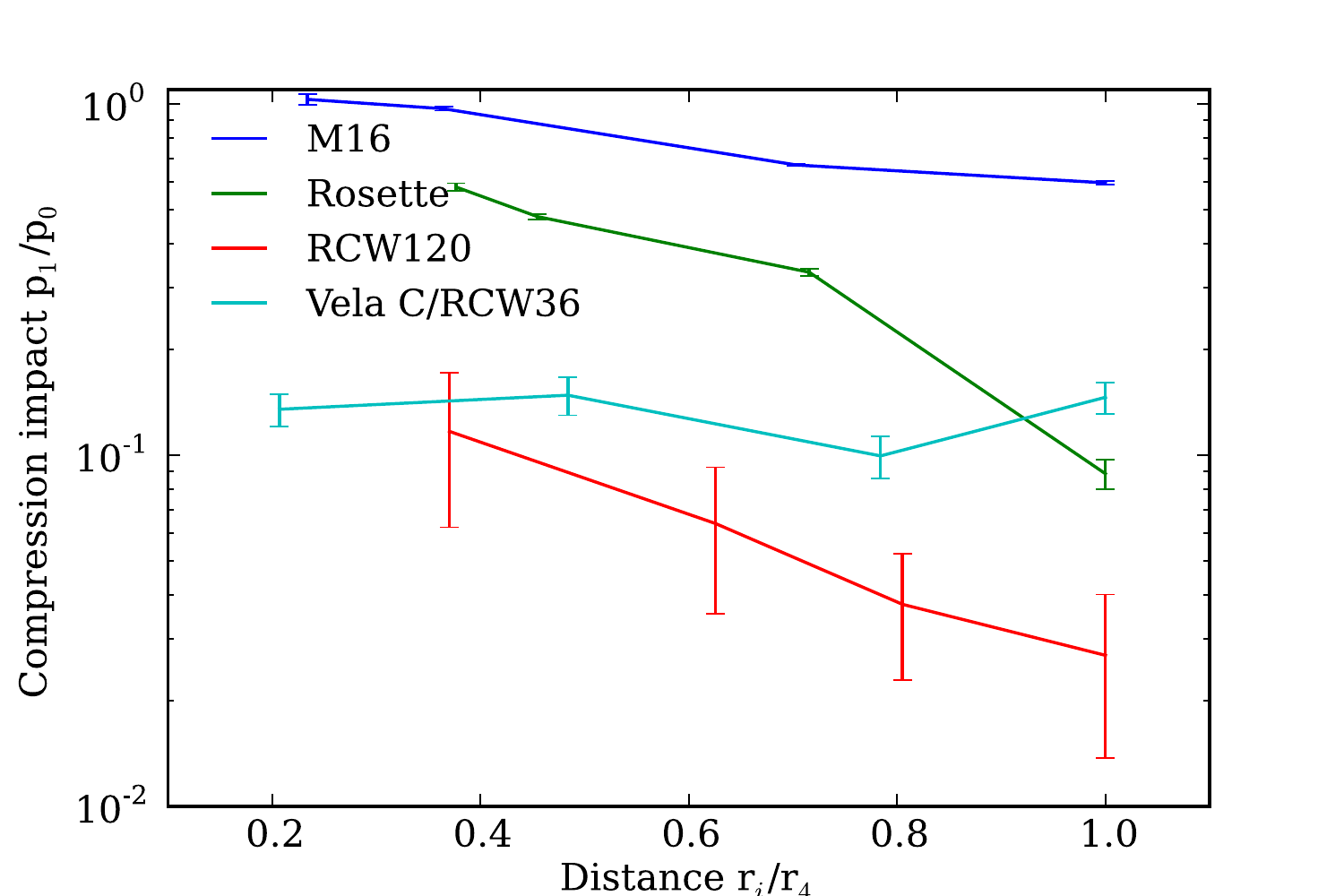}
\caption{\label{comp_impact} Radial evolution of the ratio of the number
of pixels in the compressed peak to the number of pixel in the
low-density peak $p_1/p_0$ for the four different clouds.}
\end{figure}

Based on the two-lognormal fits performed in the different regions, we
define a compression parameter as the ratio of the number of pixels in the
compressed component to the number of pixels in the low-density
component. Figure \ref{comp_impact} shows the evolution of this
parameter as a function of radius for the four different clouds. For
Rosette, M16, and for RCW 120, the compression impact
decreases with increasing radius. Therefore, the compressed material
is localized close to the ionizing sources and its relative volume
decreases when the volume of the studied region increases. 

At a first
glance, it could be argued that dense gas is present close to ionizing
sources because they were born in a dense environment formed e.g. by
gravitational collapse or colliding flows. This first possibility is
ruled out by the comparison made in M16 and Rosette between the dense
column density excess in region 1+2+3 compared to region 4
(Fig. \ref{M16_inout} and Fig. \ref{rosette_inout}). This excess does not follow a power law
as would be expected from gravitational collapse. 
In the second case, it is much more difficult to distinguish dense gas formed from
colliding flows prior to the formation of the ionization sources and
ionization compression that will happen after. However the Vela C
molecular cloud offers the possibility to show how 
the difference can be inferred. For Vela C, the dense gas does
not seem to be related to the ionizing sources as shown by
Fig. \ref{comp_impact}. The ratio of dense gas to low-density gas is
relatively constant at 15 \% in the whole region.
 From the column density map, we derive an average density in the
  Vela C molecular ridge -- assuming a thickness of the ridge of 1 pc -- of
  $10^4$ cm$^{-3}$ while in the more extended low-density medium
  ($\sim$5 pc thickness) the density is $10^2$ cm$^{-3}$.  The
  lognormal shape of the high-density component seen in regions 1+2+3
  and 1+2+3+4 suggests that the ridge is 
  formed by shock compression rather than gravitational
  instabilities. The ridge formation from shock compression caused by
  colliding flows was proposed in Vela \citep{Hill:2011ht} and also
  for example in IRDC G035.39-00.33 
  \citep{nguyenluong11b} or W43 \citep{nguyenluong13}. 
  In this case, the 
  ridge could be the result of converging flows at a density of around
  $10^2$ cm$^{-3}$ that collided at a Mach number $\sim$10 to compress
  the gas by two orders of magnitude, assuming isothermal gas. 
This estimation of the Mach number is rather close to the Mach number
expected for the cold interstellar medium ($T_c$ $\approx$ 50-100 K
for the cold neutral medium CNM) formed by the thermal instability inside the warm phase
of the interstellar medium ($T_w$ $\approx$ 8000 K for the warm
neutral medium WNM). Since
the CNM has a typical velocity that is transonic relative to the sound
speed in the WNM \citep[see][]{Audit:2010jx}, the Mach number of the
CNM is given by $M \approx \sqrt{T_w/T_c} \approx 10$. Therefore it
is possible that the Vela C molecular ridge was formed by colliding
flows whereas the dense gas in the three other regions M16, Rosette
and RCW 120 is formed by ionization compression since its volume is
inversely correlated with the distance from the ionizing sources.  

On smaller scales, we have seen that in the inner regions with smaller
pixel statistics, we can
identify power-law tails that correspond to one or two dense
condensations that have probably gravitationally collapsed. The PDF
around these condensations and their radial profiles indicate $\alpha$
exponents higher than the 
free-fall value of 2. Therefore 
they could be examples of compressed core collapse triggered by the
ionization compression. Although more studies are needed to
confirm these results,
this behavior could be the first unambiguous criterion that can make the
difference between non-triggered (free-fall) and triggered (forced-fall) core collapse around
\ion{H}{ii} regions. It could also be generalized to any type of
external compression (winds, supernovas...) whose source is identified.

\subsection{The shape of double-peaked and broadened PDFs} \label{peak}

 The regions studied here -- the Rosette and Eagle molecular
  clouds, and the RCW 36 and RCW 120 \ion{H}{ii} regions -- differ
  significantly in terms of mass (from $\sim$6$\times$10$^3$ M$_\odot$
  for RCW 120 to
  $\sim$4$\times$10$^5$ M$_\odot$ for M16), temperature and incident
  UV-flux (with a low G$_\circ$ cloud like Rosette in contrast to a
  bright cloud with a high G$_\circ$ like M16), and size (a few pc for the small \ion{H}{ii} regions up to
  $\sim$20 pc for the Rosette nebula). Nevertheless, they all show
  the same characteristic double-peaked and/or broadened PDFs.  We
  argued that it is {\sl compression caused by the expansion of the
    ionized gas} that causes a double peak in the column density PDF
  (in the case of the M16 molecular cloud) or an enlarged distribution when
  the second peak is close to the first one (in the case
  of the Rosette molecular cloud and RCW 120). As explained in the
  previous section, the Vela C molecular cloud presents a
  dense component probably formed by colliding flows that also broaden
  the PDF. The
  difference between this scenario and that of ionization compression
  can be inferred from the
  evolution of the compression parameter with increasing radius (see
  Fig.~\ref{comp_impact}).  

For Rosette, M16, and RCW 120, the compression parameter
decreases when the radius of the region on which the PDF is computed
increases. This is consistent with a compressed region localized at the
edge of the ionized gas. When the radius increases, more and more
unperturbed gas is added to the distribution.  For more extended
regions (Rosette and M16), the \ion{H}{ii} regions are
relatively large compared to the molecular cloud, and the effect and
the compression is important even on the large scale (a few
  tens of parsecs) of the cloud. In contrast, RCW 120 is relatively
small and the compression parameter is quite small compared
to the other regions (see Fig. \ref{comp_impact}).

 For these three ionization-compressed regions, we fitted the
 two-peaked and enlarged distributions 
 with two lognormals. A simple approach can qualitatively justify the
 choice of a lognormal shape for 
 the ionization-compressed component. The
first component is enlarged by the turbulence of the cloud and its lognormal
shape can be qualitatively understood
\citep[see][]{VazquezSemadeni:1994fm,Kevlahan:2009bw}. 
If a region of density $\rho_0$ is compressed by a shock of Mach
number $M_1$, the new density is then $\rho_1=\rho_0M_1^2$.  When a
second shock compresses the region, the new density is
$\rho_2=\rho_1M_2^2=\rho_0M_1^2M_2^2$ and so on... This process is
additive with respect to $\ln(\rho)$ and if it is repetitive and
random, we can expect that the logarithm of the density is a Gaussian
thanks to the central limit theorem (hence, the density should be
lognormal, see also \citet{Federrath:2010ef}). There are some limits
to this analysis 
\citep[e.g. multiple compressive forces, see][]{Hennebelle:2012dk} but
it is a straightforward way to understand 
the shape of the density distribution. 

The same idea can be used to understand the shape of the double peak
distribution for the surroundings of an \ion{H}{ii} region. As
explained in \citet{Tremblin:2012he}, there are two situations to
consider, either the ionized-gas pressure is much greater than the ram
pressure of turbulence, or turbulence dominates. In both
cases, the Mach number of the shock driven by the \ion{H}{ii} region
is well approximated by
\begin{equation}
M_\mathrm{io} = \sqrt{\zeta}\frac{c_{II}}{c_0}\left(\frac{R_s}{r_\mathrm{shell}}\right)^{3/4}
\end{equation} 
where $c_{II}$ and $c_0$ are the sound speed in the ionized and
shocked region respectively, $R_s$ is the initial Str\"omgren radius
(computed for the initial averaged density around the ionizing
source), $r_\mathrm{shell}$ the radius of the \ion{H}{ii} region at
the time considered, and $\zeta$ a parameter that is equal to two for
a D-critical front \citep{Kahn:1954tp} and one for a weakly D-type front \citep[see][for
  the detailed calculation]{Minier:2013ih}.  At the beginning of the 
development of the \ion{H}{ii} region, the ionizing front is
D-critical and $r_\mathrm{shell}=R_s$. Therefore the Mach number is of
the order of 40. Such level of turbulence is not observed in
molecular clouds, hence there is always a period of time for which
$M_\mathrm{io}\gg M_i$ where $M_i$ are the Mach numbers of the
turbulent shocks in the molecular cloud. When the shock driven by the
ionized gas passes, the gas is compressed and the log-density
increases by $\ln(M_\mathrm{io}^2)$. In an initial turbulent medium,
the same analysis as \citet{Kevlahan:2009bw} can be applied to
$\ln(\rho_0)= \ln(\rho_c)-\ln(M_\mathrm{io}^2)$. Thanks to the central
limit theorem, $\ln(\rho_0)$ is a Gaussian and $\ln(\rho_c)$ is a
Gaussian shifted by a compression factor $\ln(M_\mathrm{io}^2)$.
Therefore, if sufficient material is accumulated in the shell, two
lognormal peaks appear in the PDF. When the turbulence is high
compared to the ionized-gas pressure, $M_{io}\approx M_i$ and the term
added to the log-density appears as one shock among others in the
turbulent medium. Therefore with the central limit theorem, the
distribution of the log-density will be unchanged by the
ionization-driven shock. This is exactly what was observed in the
simulation of \citet{Tremblin:2012he}, the lognormal double-peaked PDF
appears when the ionized-gas pressure is high relative to the
turbulent ram pressure and at high turbulent level, the PDF is rather
unchanged by the ionization. This explains also why in some
  clouds, no second peak is observed but only broadening by
  compression.  A recent study of the Orion B molecular cloud
  \citep{schneider2013} showed that the PDF has a lognormal shape
  with a power-law tail but is a factor 1.5 broader than the one of a
  similar region (Aquila) that has the same Mach-number. The
  difference is that Orion B is exposed to a large-scale compression
  from a nearby OB-cluster (also seen in asymmetric column density
  profiles) while Aquila contains only a small internal \ion{H}{ii} region.
  Another example is the Auriga-California cloud \citep{Harvey:2013di}
  that shows a double-peak in the PDF. But this cloud is a low-mass
  star-forming region with only a B-star as ionizing source, so the
  total impact of the ionization front is smaller than for a cloud
  that is impacted by a whole OB-cluster. However -- as outlined above
  -- it is {\sl not} the total energy contained in the shock expansion
  that decides whether a compressed layer is formed and visible in the
  PDF, but the interplay between ionized-gas pressure and turbulent
  ram pressure. We thus do not expect a double-peaked PDF for all
  clouds associated with OB-clusters. 

  On the other hand, a
  second break in the PDF power law can be observed in some clouds and
  linked to the 
  feedback (e.g. W3, Rivera-Ingraham et al. submitted). These two breaks could
  mean that the PDFs can be fitted by 
  a lognormal for the turbulent cloud, a first power law for the
  feedback compression and a second power law for the
  influence of gravity. As mentioned in Sect.~\ref{sect_m16}, the
  inner region in 
  M16 could also be well fitted by such a model. Although our approach
  here is slightly different we think that both methods are
  complementary. In the inner region, the scales are smaller and the
  compressed shell should be rather homogeneous. In this case, the
  statistic may not be sufficient to get a proper sampling of the
  turbulent compressed peak that is seen in the turbulent-ionized
  simulations. Therefore the column-density PDF of an homogeneous
  compressed shell could lead to a flat power-law for the
  compressed component in column density, and the gravity to a power
  law with a steeper 
  slope, hence a two-step power law. On a larger scale, a sufficiently
  large range of turbulent motions will be taken into account and a
  better statistic will lead to a compressed lognormal component as
  seen in the turbulent-ionized simulations. Thus, depending on the
  scale or turbulent 
  state of the region, the compressed component may be well
  approximated by a power law or a lognormal. In both cases, the
  compression by ionization does have an impact and leads to a
  broadened PDF.

\subsection{Possible consequences of external compression for the IMF} \label{imf}

Recently, \citet{Hennebelle:2008hh,Hennebelle:2009hm,Hennebelle:2013th} proposed an analytical theory to
derive the CMF/IMF (Core/Initial Mass Function) from the PDF of
a gravo-turbulent molecular cloud. The central idea is that the shape
of the IMF can be computed from any kind of PDF assuming that the
global properties of the cloud are the determinant factor to set the
final IMF.  Therefore the typical shape of the Chabrier IMF
\citep[log-normal at low mass and a Salpeter power-law at high mass,
  see][]{Chabrier:2003ki} could be determined from 
the global properties of the gravo-turbulent cloud by computing the
gas volumes sufficiently dense to collapse gravitationally. One direct implication of the
present work is that the ionization compression affects the PDF around the
ionized gas and consequently could also affect the CMF/IMF. 

\ion{H}{ii} regions are present in a vast number of molecular 
clouds, even in low-mass star-forming regions (e.g. the Cocoon Nebula
in IC 5146, \citet{Arzoumanian:2011ho} and W40 in Aquila,
\citet{Bontemps:2010hv,Konyves:2010ff}).  
This type of feedback of high-mass stars is also thought to be one of the most important processes to disperse molecular clouds in our Galaxy
\citep[see][]{Whitworth:1979uc,Matzner:2002ii}. Considering the
typical lifetime of molecular clouds ($\approx$ 10 Myr) and the typical
time of the development of \ion{H}{ii} regions ($\approx$ 1 Myr), the
clouds should spend a large fraction of their lifetime in a state where the ionized
gas is present and compresses the cold gas to form a double-peaked or enlarged PDF
(if the conditions outlined in Sect.~\ref{peak} are fulfilled).
It is therefore unlikely that the dispersion of the molecular gas by
ionization is sufficiently rapid to safely ignore its impact on the
PDF of the gas that will form stars. It is also unlikely that other feedback processes
such as radiation pressure and stellar winds are sufficiently
rapid to ignore the compression phase that will disperse the gas to
get a gas-free cluster with a given IMF. 

Multiple \ion{H}{ii} regions may interact, but it is
rather unlikely that the molecular clouds presented here would produce
enough distinct \ion{H}{ii} regions to restore a
lognormal shape with the statistical argument of
\citet{Kevlahan:2009bw}. In all observations presented here, only one
or two 
\ion{H}{ii} regions are present in the cloud. However it is
sufficient to enlarge the PDF with a rather small initial Mach number for the
cloud. For example, the width of the PDF of the ionized Mach-1 simulation
in \citet{Tremblin:2012he} is equivalent to the width of the Mach-4
simulation. Therefore we propose that the feedback processes such as the ionization
could account for the relatively large PDF that are needed in
\citet{Hennebelle:2008hh} to have a good agreement with the computed
and observed IMF. Typically compressed gas could help forming
compressed dense and massive cores as seen in Rosette, RCW 120 and RCW
36 in the present study but also small
and dense cores leading to the formation of brown dwarfs
\citep[see][]{Padoan:2004ez}. However 
high-resolution on large scales is needed to be able to have a full
statistic of the cores around \ion{H}{ii} regions.
Since other processes could also enlarge the PDF
\citep[e.g. the equation of state and variations among the core
  properties, see][]{Hennebelle:2009hm,Chabrier:2010hk}, the 
relative importance of the physical phenomena leading to these large
PDF in molecular clouds remains a relatively open question although
the present work observationally supports the role of ionization. 

\subsection{Pre-existing versus triggered dense structures}

A number of recent {\it Herschel} studies showed the importance of gas
accretion by filaments, both for low-mass
\citep[e.g.][]{Arzoumanian:2011ho,Peretto:2012bu,Palmeirim:2013da} and
high-mass \citep[e.g.][]{Schneider:2012ds,Hennemann:2012dp,nguyenluong13} star-forming
regions, consistent with the accretion-driven turbulence scenario
\citep{Klessen:2010jh}. Filaments may form as a result of the
dissipation of large-scale turbulence
\citep{Elmegreen:2004jb,Federrath:2013vo} or by gravity 
\citep{Bonnell:2008df}. Because pre-stellar cores are found inside
dense self-gravitating filaments
\citep[e.g.][]{Andre:2010ka}, the general view is
that the accumulation of matter 
leads to the formation of self-gravitating cores by gravitational
fragmentation. These results
suggest that the CMF from the filamentary structures 
resembles the stellar IMF which is consistent with the gravo-turbulent
scenario \citep[see][]{Motte:1998un,Andre:2010ka}. However, because pre-stellar cores are
primarily found inside dense self-gravitating filaments,  
the {\it Herschel} results on nearby clouds 
suggest that the peak of the pre-stellar CMF may result from the pure
gravitational fragmentation of filaments, independently of the cloud
PDF \citep[see][]{Andre:2010ka,Andre:2011jp}. 
A detailed analysis of the
data is therefore needed to fully 
characterize the CMF/IMF relationship (e.g. environmental effects on
the M$_*$/M$_\mathrm{core}$ conversion) and also to determine precisely
the role of the turbulence, gravity, and feedback in the PDF/CMF relationship. In
addition to individual core collapse, large-scale 
gravitational collapse across a length scale of several parsecs
probably plays an important role as well, in particular for the formation of
high-mass stars. \citet{Schneider:2010kx} showed that the global
collapse of the DR21 filament {\sl and} mass input by filaments would
explain the formation of the proto-OB cluster region DR21(OH). A
scenario which was confirmed by {\sl Herschel} observations
\citep{Hennemann:2012dp} and studies of other regions
\citep[e.g.][]{GalvanMadrid:2010jw,Peretto:2013kt}.  

The effect of the ionizing compression on these pre-existing massive filaments
and gravitationally-bound clumps could be marginal since these
structures are already quite dense. One could expect, for example, that
the ionization only erodes the structures and that they keep radial
profiles dominated by gravity, i.e. $\propto r^{-\alpha}$ with $\alpha$
between 1.5 and 2.
However this effect is still
relatively unknown. Furthermore the relative importance of triggered and pre-existing dense
structures is highly debated. While statistical observations
\citep{Thompson:2012gn} show that massive young stellar objects are
found preferentially at the edge of \ion{H}{ii} regions, numerical
studies \citep{Dale:2013da} shows that it is very difficult to
separate the pre-existing and triggered star formation even at the
edge of ionized regions. Observationally, only on the
closest regions like RCW 36 in Vela C, a 
detailed study of the link between compression and the dense
clumps is possible  thanks to velocity
information (Minier et al. 2013). In this region it was shown that the dense cores are
associated with the dense shell around the ionized gas and were
formed by the compression effect caused by ionization before becoming gravitationally
unstable. Furthermore, the forced-fall collapses with steep radial
profiles confirm this analysis.
Based on this work and the large-scale
compression highlighted in the present paper, we postulate that the
feedback helps to form denser cores and consequently a possible broader CMF. A
systematic comparison of the shape of the CMF between isolated regions
and regions influenced by the feedback is needed to confirm this
hypothesis. Especially large-scale mapping at a very high angular resolution
will help computing PDFs with a large pixel statistic. These PDFs and
the radial profiles of the cores
could be used to distinguish between free-fall and compressed core
collapse with the criterion used in the present paper. 
Since most of the high-mass star-forming regions are
located at more than 1000 pc from the Solar System, this systematic comparison is
out of reach with the present observatories and could only be done
thanks to a high-resolution (typically 1$''$ at 100 $\mu$m),
 and wide-field submillimeter instrument with high sensitivity. 

\section{Conclusions and perspectives}

Thanks to the recent {\it Herschel} observations, we performed an analysis
of the column-density structure of four molecular clouds 
around \ion{H}{ii} regions, namely M16, the Rosette and Vela C molecular
cloud, and the RCW 120 \ion{H}{ii} region. The compression induced by the
expansion 
of the ionized gas leads to PDFs that are enlarged (as in Rosette, and
the RCW 120) or double peaked (as in M16). The form
is determined by the relative importance between the initial
turbulence of the molecular cloud and the compression. This result was
also predicted by turbulent-ionized numerical simulations. If the PDF
of the molecular cloud is a log-normal, the peak induced by the
compression can be qualitatively understood to be also a log-normal,
shifted to higher densities. The difference between a dense lognormal
induced by colliding flows and the ionization compression can be made
by studying the evolution of the PDFs around the ionizing sources
(i.e. Vela C versus other regions in the present paper).
In addition to that, a power-law tail can
also be identified for the highest column density range and can be
attributed to the effect of the gravitational force. Especially there
are good candidates of single core-collapse in some of the regions
that can be identified thanks to the PDFs and localized by looking at
the spatial distribution of the contributing pixels. Furthermore a criterion based on the
power law and the radial profiles can be derived to distinguish between free-fall and
compressed core collapse. This criterion  could be used to unambiguously
make the difference between triggered and pre-existing star
formation (free-fall core collapse with a radial profile $\propto
r^{-2}$ and forced-fall core collapse with a steeper profile). In this
paper, the triggering mechanism is the ionization but this criterion
could also be generalized to other types of external compression. 

The enlarged PDFs are important as they may lead to an IMF that matches
the observations while keeping a relatively low Mach number for the
initial cloud. Since ionization is thought to be one of the primary
effects for the dispersion of molecular clouds, its compression effect
during the lifetime of the cloud could be a key ingredient to explain
the relatively large PDF that are needed by a gravo-turbulent scenario
to get a reasonable IMF. A detailed theoretical study of the passage between these
ionized-enlarged PDFs to the final IMF has to be done in order to confirm this
link. Especially such a study could tell what is the relative
importance of the different effects, the compression
observed in this paper, the equation of state or the variations among
the core properties. Furthermore a detailed observational study of the
cores around \ion{H}{ii} regions and in feedback-isolated regions is
needed to complete our understanding of the effect of the compression
on the CMF. Since the PDF is observed to be affected by the
ionization compression, such a study could also help to determine the nature of the
link between the PDF and the CMF. A submillimetre space observatory with
high sensitivity, wide field, and very high resolution will bring the
data needed to fulfill this task.

%
%

\begin{acknowledgements}
SPIRE has been developed by a consortium of institutes led by Cardiff
Univ. (UK) and including: Univ. Lethbridge (Canada); NAOC (China);
CEA, LAM (France); IFSI, Univ. Padua (Italy); IAC (Spain); Stockholm
Observatory (Sweden); Imperial College London, RAL, UCL-MSSL, UKATC,
Univ. Sussex (UK); and Caltech, JPL, NHSC, Univ. Colorado (USA). This
development has been supported by national funding agencies: CSA
(Canada); NAOC (China); CEA, CNES, CNRS (France); ASI (Italy); MCINN
(Spain); SNSB (Sweden); STFC, UKSA (UK); and NASA (USA). 
PACS has been developed by a consortium of institutes led by MPE
(Germany) and including UVIE (Austria); KU Leuven, CSL, IMEC
(Belgium); CEA, LAM (France); MPIA (Germany); INAF-IFSI/OAA/OAP/OAT,
LENS, SISSA (Italy); IAC (Spain). This development has been supported
by the funding agencies BMVIT (Austria), ESA-PRODEX (Belgium),
CEA/CNES (France), DLR (Germany), ASI/INAF (Italy), and CICYT/MCYT
(Spain). Part of this work was supported by the  ANR-11-BS56-010  
project ``STARFICH''.  
\end{acknowledgements}

%
%

\appendix
\section{Concentric PDFs of Rosette with a regular increase in radius}
We show in this appendix the concentric PDFs of Rosette for another
choice of radius, regularly spaced between the inner and outer
disks. The shape of the PDF in region 1+2, 1+2+3, and 1+2+3+4 is
relatively similar. They all include the central star-forming region,
therefore this region is indeed the important structure that shapes
the power-law tail of the three distributions. The compression
parameter p$_1$/p$_0$ still decreases from 0.58 in region 1 down to 0.09 in
region 1+2+3+4.
\begin{figure}[!h]
\centering
\includegraphics[trim=0 3cm 2cm 0,width=\linewidth]{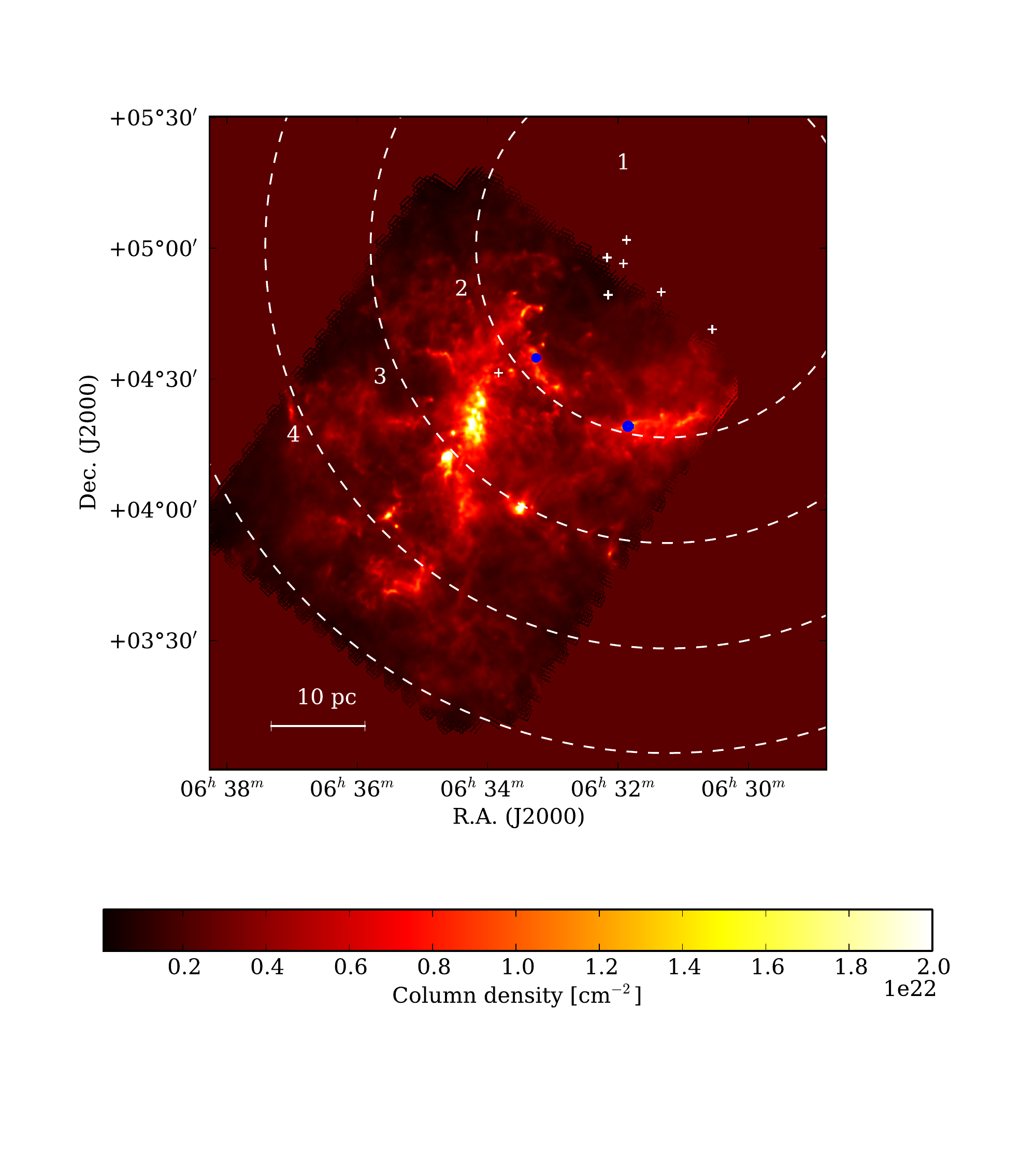}
\caption{\label{rosette_N_pdf} Rosette {\it Herschel} column density
  map \citep{Schneider:2012ds}. The circles indicate the different
  regions used for the PDFs. The white crosses indicate the position of
  the main ionizing sources, i.e. the most massive OB-stars from the
  NGC2244 cluster. The radius of the disks are regularly spaced
  between the inner and outer disks.}
\end{figure}
\begin{figure}[!h]
\centering
\includegraphics[trim=0 0 1cm 0,width=\linewidth]{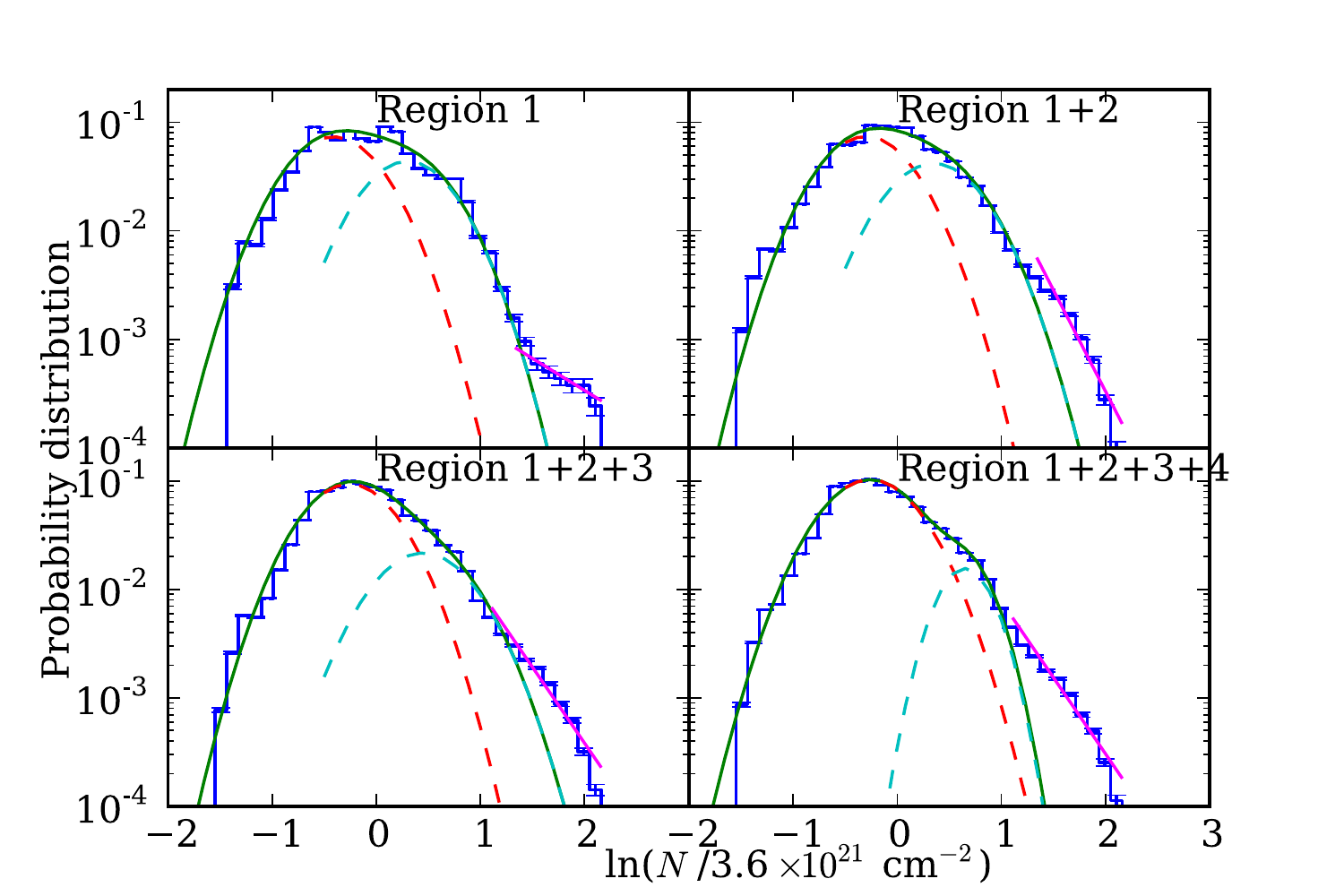}
\captionof{figure}{\label{rosette_pdf}  Rosette PDFs of the four regions
  indicated in Fig. \ref{rosette_N_pdf}. The multi-component fit is done using
  lognormal distributions (see Eq. \ref{eq_fit}) and a power law at
  high column densities (see Eq. \ref{eq_pl}).}
\end{figure}

\bibliographystyle{aa}
\bibliography{main.bib}

\providecommand{\SortNoop}[1]{}
\begin{thebibliography}{104}
\expandafter\ifx\csname natexlab\endcsname\relax\def\natexlab#1{#1}\fi

\bibitem[{Allen {et~al.}(1999)Allen, Burton, Ryder, Ashley, \&
  Storey}]{Allen:1999cl}
Allen, L.~E., Burton, M.~G., Ryder, S.~D., Ashley, M. C.~B., \& Storey, J.
  W.~V. 1999, MNRAS, 304, 98

\bibitem[{Anderson {et~al.}(2012)Anderson, Zavagno, Deharveng, Abergel, Motte,
  Andr{\'e}, Bernard, Bontemps, Hennemann, Hill, Rod{\'o}n, Roussel, \&
  Russeil}]{Anderson:2012jy}
Anderson, L.~D., Zavagno, A., Deharveng, L., {et~al.} 2012, A{\&}A, 542, 10

\bibitem[{Anderson {et~al.}(2010)Anderson, Zavagno, Rod{\'o}n, Russeil,
  Abergel, Ade, Andr{\'e}, Arab, Baluteau, Bernard, Blagrave, Bontemps,
  Boulanger, Cohen, Compi{\`e}gne, Cox, Dartois, Davis, Emery, Fulton, Gry,
  Habart, Huang, Joblin, Jones, Kirk, Lagache, Lim, Madden, Makiwa, Martin,
  Miville-Desch{\^e}nes, Molinari, Moseley, Motte, Naylor, Okumura,
  Pinheiro~Gon{\c c}alves, Polehampton, Saraceno, Sauvage, Sidher, Spencer,
  Swinyard, Ward-Thompson, \& White}]{Anderson:2010dp}
Anderson, L.~D., Zavagno, A., Rod{\'o}n, J.~A., {et~al.} 2010, A{\&}A, 518, L99

\bibitem[{Andr{\'e} {et~al.}(2010)Andr{\'e}, Men'shchikov, Bontemps,
  K{\"o}nyves, Motte, Schneider, Didelon, Minier, Saraceno, Ward-Thompson,
  Di~Francesco, White, Molinari, Testi, Abergel, Griffin, Henning, Royer,
  Mer{\'\i}n, Vavrek, Attard, Arzoumanian, Wilson, Ade, Aussel, Baluteau,
  Benedettini, Bernard, Blommaert, Cambr{\'e}sy, Cox, Di~Giorgio, Hargrave,
  Hennemann, Huang, Kirk, Krause, Launhardt, Leeks, Le~Pennec, Li, Martin,
  Maury, Olofsson, Omont, Peretto, Pezzuto, Prusti, Roussel, Russeil, Sauvage,
  Sibthorpe, Sicilia-Aguilar, Spinoglio, Waelkens, Woodcraft, \&
  Zavagno}]{Andre:2010ka}
Andr{\'e}, P., Men'shchikov, A., Bontemps, S., {et~al.} 2010, A{\&}A, 518, L102

\bibitem[{Andr{\'e} {et~al.}(2011)Andr{\'e}, Men'shchikov, K{\"o}nyves, \&
  Arzoumanian}]{Andre:2011jp}
Andr{\'e}, P., Men'shchikov, A., K{\"o}nyves, V., \& Arzoumanian, D. 2011,
  Computational Star Formation, 270, 255

\bibitem[{Arzoumanian {et~al.}(2011)Arzoumanian, Andr{\'e}, Didelon,
  K{\"o}nyves, Schneider, Men'shchikov, Sousbie, Zavagno, Bontemps,
  Di~Francesco, Griffin, Hennemann, Hill, Kirk, Martin, Minier, Molinari,
  Motte, Peretto, Pezzuto, Spinoglio, Ward-Thompson, White, \&
  Wilson}]{Arzoumanian:2011ho}
Arzoumanian, D., Andr{\'e}, P., Didelon, P., {et~al.} 2011, A{\&}A, 529, L6

\bibitem[{Audit \& Hennebelle(2010)}]{Audit:2010jx}
Audit, E. \& Hennebelle, P. 2010, A{\&}A, 511, 76

\bibitem[{Baba {et~al.}(2004)Baba, Nagata, Nagayama, Nagashima, Kato, Kurita,
  Sato, Nakajima, Tamura, Nakaya, \& Sugitani}]{Baba:2004gm}
Baba, D., Nagata, T., Nagayama, T., {et~al.} 2004, ApJ, 614, 818

\bibitem[{Bertoldi(1989)}]{Bertoldi:1989bq}
Bertoldi, F. 1989, ApJ, 346, 735

\bibitem[{Bevington \& Robinson(2003)}]{Bevington:2003tc}
Bevington, P.~R. \& Robinson, D.~K. 2003, Data reduction and error analysis for
  the physical sciences

\bibitem[{Bisbas {et~al.}(2011)Bisbas, W{\"u}nsch, Whitworth, Hubber, \&
  Walch}]{Bisbas:2011kg}
Bisbas, T.~G., W{\"u}nsch, R., Whitworth, A.~P., Hubber, D.~A., \& Walch, S.
  2011, ApJ, 736, 142

\bibitem[{Bohlin {et~al.}(1978)Bohlin, Savage, \& Drake}]{Bohlin:1978dw}
Bohlin, R.~C., Savage, B.~D., \& Drake, J.~F. 1978, ApJ, 224, 132

\bibitem[{Bonatto {et~al.}(2006)Bonatto, Santos, \& Bica}]{Bonatto:2006kb}
Bonatto, C., Santos, J. F. C.~J., \& Bica, E. 2006, A{\&}A, 445, 567

\bibitem[{Bonnell {et~al.}(2008)Bonnell, Clark, \& Bate}]{Bonnell:2008df}
Bonnell, I.~A., Clark, P., \& Bate, M.~R. 2008, MNRAS, 389, 1556

\bibitem[{Bontemps {et~al.}(2010)Bontemps, Andr{\'e}, K{\"o}nyves,
  Men'shchikov, Schneider, Maury, Peretto, Arzoumanian, Attard, Motte, Minier,
  Didelon, Saraceno, Abergel, Baluteau, Bernard, Cambr{\'e}sy, Cox,
  Di~Francesco, di~Giorgo, Griffin, Hargrave, Huang, Kirk, Li, Martin,
  Mer{\'\i}n, Molinari, Olofsson, Pezzuto, Prusti, Roussel, Russeil, Sauvage,
  Sibthorpe, Spinoglio, Testi, Vavrek, Ward-Thompson, White, Wilson, Woodcraft,
  \& Zavagno}]{Bontemps:2010hv}
Bontemps, S., Andr{\'e}, P., K{\"o}nyves, V., {et~al.} 2010, A{\&}A, 518, L85

\bibitem[{Caswell \& Haynes(1987)}]{Caswell:1987tc}
Caswell, J.~L. \& Haynes, R.~F. 1987, A{\&}A, 171, 261

\bibitem[{Chabrier(2003)}]{Chabrier:2003ki}
Chabrier, G. 2003, PASP, 115, 763

\bibitem[{Chabrier \& Hennebelle(2010)}]{Chabrier:2010hk}
Chabrier, G. \& Hennebelle, P. 2010, ApJ, 725, L79

\bibitem[{Cox {et~al.}(1990)Cox, Deharveng, \& Leene}]{Cox:1990ub}
Cox, P., Deharveng, L., \& Leene, A. 1990, A{\&}A, 230, 181

\bibitem[{Dale \& Bonnell(2011)}]{Dale:2011ct}
Dale, J.~E. \& Bonnell, I. 2011, MNRAS, 414, 321

\bibitem[{Dale {et~al.}(2007)Dale, Clark, \& Bonnell}]{Dale:2007cr}
Dale, J.~E., Clark, P.~C., \& Bonnell, I.~A. 2007, MNRAS, 377, 535

\bibitem[{Dale {et~al.}(2013)Dale, Ercolano, \& Bonnell}]{Dale:2013da}
Dale, J.~E., Ercolano, B., \& Bonnell, I.~A. 2013, Monthly Notices of the Royal
  Astronomical Society, 431, 1062

\bibitem[{Deharveng {et~al.}(2009)Deharveng, Zavagno, Schuller, Caplan,
  Pomar{\`e}s, \& De~Breuck}]{Deharveng:2009kd}
Deharveng, L., Zavagno, A., Schuller, F., {et~al.} 2009, A{\&}A, 496, 177

\bibitem[{Di~Francesco {et~al.}(2010)Di~Francesco, Sadavoy, Motte, Schneider,
  Hennemann, Csengeri, Bontemps, Balog, Zavagno, Andr{\'e}, Saraceno, Griffin,
  Men'shchikov, Abergel, Baluteau, Bernard, Cox, Deharveng, Didelon,
  di~Giorgio, Hargrave, Huang, Kirk, Leeks, Li, Marston, Martin, Minier,
  Molinari, Olofsson, Persi, Pezzuto, Russeil, Sauvage, Sibthorpe, Spinoglio,
  Testi, Teyssier, Vavrek, Ward-Thompson, White, Wilson, \&
  Woodcraft}]{DiFrancesco:2010ef}
Di~Francesco, J., Sadavoy, S., Motte, F., {et~al.} 2010, A{\&}A, 518, L91

\bibitem[{Ellerbroek {et~al.}(2013)Ellerbroek, Bik, Kaper, Maaskant, Paalvast,
  Tramper, Sana, Waters, \& Balog}]{Ellerbroek:2013tv}
Ellerbroek, L.~E., Bik, A., Kaper, L., {et~al.} 2013, arXiv, 3238

\bibitem[{Elmegreen \& Lada(1977)}]{Elmegreen:1977iq}
Elmegreen, B.~G. \& Lada, C.~J. 1977, ApJ, 214, 725

\bibitem[{Elmegreen \& Scalo(2004)}]{Elmegreen:2004jb}
Elmegreen, B.~G. \& Scalo, J. 2004, Annual Review of Astronomy
  {\&}Astrophysics, 42, 211

\bibitem[{Federrath(2013)}]{Federrath:2013vo}
Federrath, C. 2013, arXiv, 3989

\bibitem[{Federrath \& Klessen(2012)}]{Federrath:2012em}
Federrath, C. \& Klessen, R.~S. 2012, ApJ, 761, 156

\bibitem[{Federrath \& Klessen(2013)}]{Federrath:2013ip}
Federrath, C. \& Klessen, R.~S. 2013, ApJ, 763, 51

\bibitem[{Federrath {et~al.}(2008)Federrath, Klessen, \&
  Schmidt}]{Federrath:2008ey}
Federrath, C., Klessen, R.~S., \& Schmidt, W. 2008, ApJ, 688, L79

\bibitem[{Federrath {et~al.}(2010)Federrath, Roman-Duval, Klessen, Schmidt, \&
  Mac~Low}]{Federrath:2010ef}
Federrath, C., Roman-Duval, J., Klessen, R.~S., Schmidt, W., \& Mac~Low, M.-M.
  2010, A{\&}A, 512, 81

\bibitem[{Flagey {et~al.}(2011)Flagey, Boulanger, Noriega-Crespo, Paladini,
  Montmerle, Carey, Gagn{\'e}, \& Shenoy}]{Flagey:2011jr}
Flagey, N., Boulanger, F., Noriega-Crespo, A., {et~al.} 2011, A{\&}A, 531, 51

\bibitem[{Galv{\'a}n-Madrid {et~al.}(2010)Galv{\'a}n-Madrid, Zhang, Keto, Ho,
  Zapata, Rodr{\'\i}guez, Pineda, \&
  V{\'a}zquez-Semadeni}]{GalvanMadrid:2010jw}
Galv{\'a}n-Madrid, R., Zhang, Q., Keto, E., {et~al.} 2010, ApJ, 725, 17

\bibitem[{Gritschneder {et~al.}(2010)Gritschneder, Burkert, Naab, \&
  Walch}]{Gritschneder:2010du}
Gritschneder, M., Burkert, A., Naab, T., \& Walch, S. 2010, ApJ, 723, 971

\bibitem[{Harvey {et~al.}(2013)Harvey, Fallscheer, Ginsburg, Terebey,
  Andr{\'e}, Bourke, Di~Francesco, K{\"o}nyves, Matthews, \&
  Peterson}]{Harvey:2013di}
Harvey, P.~M., Fallscheer, C., Ginsburg, A., {et~al.} 2013, The Astrophysical
  Journal, 764, 133

\bibitem[{Haworth \& Harries(2011)}]{Haworth:2011gv}
Haworth, T.~J. \& Harries, T.~J. 2011, MNRAS, 420, 562

\bibitem[{Hennebelle \& Chabrier(2008)}]{Hennebelle:2008hh}
Hennebelle, P. \& Chabrier, G. 2008, ApJ, 684, 395

\bibitem[{Hennebelle \& Chabrier(2009)}]{Hennebelle:2009hm}
Hennebelle, P. \& Chabrier, G. 2009, ApJ, 702, 1428

\bibitem[{Hennebelle \& Chabrier(2013)}]{Hennebelle:2013th}
Hennebelle, P. \& Chabrier, G. 2013, arXiv, 6637

\bibitem[{Hennebelle \& Falgarone(2012)}]{Hennebelle:2012dk}
Hennebelle, P. \& Falgarone, E. 2012, The Astronomy and Astrophysics Review,
  20, 55

\bibitem[{Hennemann {et~al.}(2010)Hennemann, Motte, Bontemps, Schneider,
  Csengeri, Balog, Di~Francesco, Zavagno, Andr{\'e}, Men'shchikov, Abergel,
  Ali, Baluteau, Bernard, Cox, Didelon, di~Giorgio, Griffin, Hargrave, Hill,
  Horeau, Huang, Kirk, Leeks, Li, Marston, Martin, Molinari, Nguyen-Luong,
  Olofsson, Persi, Pezzuto, Russeil, Saraceno, Sauvage, Sibthorpe, Spinoglio,
  Testi, Ward-Thompson, White, Wilson, \& Woodcraft}]{Hennemann:2010hp}
Hennemann, M., Motte, F., Bontemps, S., {et~al.} 2010, A{\&}A, 518, L84

\bibitem[{Hennemann {et~al.}(2012)Hennemann, Motte, Schneider, Didelon, Hill,
  Arzoumanian, Bontemps, Csengeri, Andr{\'e}, K{\"o}nyves, Louvet, Marston,
  Men'shchikov, Minier, Nguyen-Luong, Palmeirim, Peretto, Sauvage, Zavagno,
  Anderson, Bernard, Di~Francesco, Elia, Li, Martin, Molinari, Pezzuto,
  Russeil, Rygl, Schisano, Spinoglio, Sousbie, Ward-Thompson, \&
  White}]{Hennemann:2012dp}
Hennemann, M., Motte, F., Schneider, N., {et~al.} 2012, A{\&}A, 543, L3

\bibitem[{Hester {et~al.}(1996)Hester, Scowen, Sankrit, Lauer, Ajhar, Baum,
  Code, Currie, Danielson, Ewald, Faber, Grillmair, Groth, Holtzman, Hunter,
  Kristian, Light, Lynds, Monet, O'Neil, Shaya, Seidelmann, \&
  Westphal}]{Hester:1996ir}
Hester, J.~J., Scowen, P.~A., Sankrit, R., {et~al.} 1996, Astronomical Journal
  v.111, 111, 2349

\bibitem[{Heyer {et~al.}(2006)Heyer, Williams, \& Brunt}]{Heyer:2006hv}
Heyer, M.~H., Williams, J.~P., \& Brunt, C.~M. 2006, ApJ, 643, 956

\bibitem[{Hildebrand(1983)}]{Hildebrand:1983tm}
Hildebrand, R.~H. 1983, Quarterly Journal of the Royal Astronomical Society,
  24, 267

\bibitem[{Hill {et~al.}(2012{\natexlab{a}})Hill, Andr{\'e}, Arzoumanian, Motte,
  Minier, Men'shchikov, Didelon, Hennemann, K{\"o}nyves, Nguyen-Luong,
  Palmeirim, Peretto, Schneider, Bontemps, Louvet, Elia, Giannini, Rev{\'e}ret,
  Le~Pennec, Rodriguez, Boulade, Doumayrou, Dubreuil, Gallais, Lortholary,
  Martignac, Talvard, \& De~Breuck}]{Hill:2012ip}
Hill, T., Andr{\'e}, P., Arzoumanian, D., {et~al.} 2012{\natexlab{a}}, A{\&}A,
  548, L6

\bibitem[{Hill {et~al.}(2011)Hill, Motte, Didelon, Bontemps, Minier, Hennemann,
  Schneider, Andr{\'e}, Men'shchikov, Anderson, Arzoumanian, Bernard,
  Di~Francesco, Elia, Giannini, Griffin, K{\"o}nyves, Kirk, Marston, Martin,
  Molinari, Nguyen-Luong, Peretto, Pezzuto, Roussel, Sauvage, Sousbie, Testi,
  Ward-Thompson, White, Wilson, \& Zavagno}]{Hill:2011ht}
Hill, T., Motte, F., Didelon, P., {et~al.} 2011, A{\&}A, 533, 94

\bibitem[{Hill {et~al.}(2012{\natexlab{b}})Hill, Motte, Didelon, White,
  Marston, Nguyen-Luong, Bontemps, Andr{\'e}, Schneider, Hennemann, Sauvage,
  Di~Francesco, Minier, Anderson, Bernard, Elia, Griffin, Li, Peretto, Pezzuto,
  Polychroni, Roussel, Rygl, Schisano, Sousbie, Testi, Thompson, \&
  Zavagno}]{Hill:2012jb}
Hill, T., Motte, F., Didelon, P., {et~al.} 2012{\natexlab{b}}, A{\&}A, 542, 114

\bibitem[{Kahn(1954)}]{Kahn:1954tp}
Kahn, F.~D. 1954, Bulletin of the Astronomical Institutes of the Netherlands,
  12, 187

\bibitem[{Kevlahan \& Pudritz(2009)}]{Kevlahan:2009bw}
Kevlahan, N. \& Pudritz, R.~E. 2009, ApJ, 702, 39

\bibitem[{Klessen {et~al.}(2000)Klessen, Heitsch, \& Mac~Low}]{Klessen:2000ca}
Klessen, R.~S., Heitsch, F., \& Mac~Low, M.-M. 2000, ApJ, 535, 887

\bibitem[{Klessen \& Hennebelle(2010)}]{Klessen:2010jh}
Klessen, R.~S. \& Hennebelle, P. 2010, A{\&}A, 520, 17

\bibitem[{K{\"o}nyves {et~al.}(2010)K{\"o}nyves, Andr{\'e}, Men'shchikov,
  Schneider, Arzoumanian, Bontemps, Attard, Motte, Didelon, Maury, Abergel,
  Ali, Baluteau, Bernard, Cambr{\'e}sy, Cox, Di~Francesco, di~Giorgio, Griffin,
  Hargrave, Huang, Kirk, Li, Martin, Minier, Molinari, Olofsson, Pezzuto,
  Russeil, Roussel, Saraceno, Sauvage, Sibthorpe, Spinoglio, Testi,
  Ward-Thompson, White, Wilson, Woodcraft, \& Zavagno}]{Konyves:2010ff}
K{\"o}nyves, V., Andr{\'e}, P., Men'shchikov, A., {et~al.} 2010, A{\&}A, 518,
  L106

\bibitem[{Kritsuk {et~al.}(2007)Kritsuk, Norman, Padoan, \&
  Wagner}]{Kritsuk:2007gn}
Kritsuk, A.~G., Norman, M.~L., Padoan, P., \& Wagner, R. 2007, ApJ, 665, 416

\bibitem[{Kritsuk {et~al.}(2011)Kritsuk, Norman, \& Wagner}]{Kritsuk:2011hw}
Kritsuk, A.~G., Norman, M.~L., \& Wagner, R. 2011, ApJ, 727, L20

\bibitem[{Lefloch \& Lazareff(1994)}]{Lefloch:1994ts}
Lefloch, B. \& Lazareff, B. 1994, A{\&}A, 289, 559

\bibitem[{Lombardi {et~al.}(2006)Lombardi, Alves, \& Lada}]{Lombardi:2006fx}
Lombardi, M., Alves, J., \& Lada, C.~J. 2006, A{\&}A, 454, 781

\bibitem[{Mackey \& Lim(2010)}]{Mackey:2010cv}
Mackey, J. \& Lim, A.~J. 2010, MNRAS, 403, 714

\bibitem[{Martins {et~al.}(2010)Martins, Pomar{\`e}s, Deharveng, Zavagno, \&
  Bouret}]{Martins:2010jw}
Martins, F., Pomar{\`e}s, M., Deharveng, L., Zavagno, A., \& Bouret, J.~C.
  2010, A{\&}A, 510, 32

\bibitem[{Matzner(2002)}]{Matzner:2002ii}
Matzner, C.~D. 2002, ApJ, 566, 302

\bibitem[{Miao {et~al.}(2006)Miao, White, Nelson, Thompson, \&
  Morgan}]{Miao:2006bx}
Miao, J., White, G.~J., Nelson, R., Thompson, M., \& Morgan, L. 2006, MNRAS,
  369, 143

\bibitem[{Miao {et~al.}(2009)Miao, White, Thompson, \& Nelson}]{Miao:2009ds}
Miao, J., White, G.~J., Thompson, M.~A., \& Nelson, R.~P. 2009, ApJ, 692, 382

\bibitem[{Minier {et~al.}(2013)Minier, Tremblin, Hill, Motte, Andr{\'e}, Lo,
  Schneider, Audit, White, Hennemann, Cunningham, Deharveng, Didelon,
  Di~Francesco, Elia, Giannini, Nguyen-Luong, Pezzuto, Rygl, Spinoglio,
  Ward-Thompson, \& Zavagno}]{Minier:2013ih}
Minier, V., Tremblin, P., Hill, T., {et~al.} 2013, A{\&}A, 550, 50

\bibitem[{Molina {et~al.}(2012)Molina, Glover, Federrath, \&
  Klessen}]{Molina:2012iv}
Molina, F.~Z., Glover, S. C.~O., Federrath, C., \& Klessen, R.~S. 2012, Monthly
  Notices of the Royal Astronomical Society, 423, 2680

\bibitem[{Motte {et~al.}(1998)Motte, Andr{\'e}, \& Neri}]{Motte:1998un}
Motte, F., Andr{\'e}, P., \& Neri, R. 1998, A{\&}A, 336, 150

\bibitem[{Motte {et~al.}(2012)Motte, Bontemps, Hennemann, Nguyen-Luong,
  Schneider, Didelon, \& Zavagno}]{Motte:2012ut}
Motte, F., Bontemps, S., Hennemann, M., {et~al.} 2012, in SF2A-2012:
  Proceedings of the Annual meeting of the French Society of Astronomy and
  Astrophysics. Eds.: S. Boissier, 45--50

\bibitem[{Motte {et~al.}(2010)Motte, Zavagno, Bontemps, Schneider, Hennemann,
  Di~Francesco, Andr{\'e}, Saraceno, Griffin, Marston, Ward-Thompson, White,
  Minier, Men'shchikov, Hill, Abergel, Anderson, Aussel, Balog, Baluteau,
  Bernard, Cox, Csengeri, Deharveng, Didelon, di~Giorgio, Hargrave, Huang,
  Kirk, Leeks, Li, Martin, Molinari, Nguyen-Luong, Olofsson, Persi, Peretto,
  Pezzuto, Roussel, Russeil, Sadavoy, Sauvage, Sibthorpe, Spinoglio, Testi,
  Teyssier, Vavrek, Wilson, \& Woodcraft}]{Motte:2010fy}
Motte, F., Zavagno, A., Bontemps, S., {et~al.} 2010, A{\&}A, 518, L77

\bibitem[{Murphy \& May(1991)}]{Murphy:1991ww}
Murphy, D.~C. \& May, J. 1991, A{\&}A, 247, 202

\bibitem[{{Nguyen Luong} {et~al.}(2013){Nguyen Luong}, {Motte}, {Carlhoff},
  {Louvet}, {Lesaffre}, {Schilke}, {Hill}, {Hennemann}, {Gusdorf}, {Schneider},
  {Bontemps}, {Duarte-Cabral}, {Menten}, {Martin}, {Wyrowski}, {Bendo},
  {Roussel}, {Bernard}, {Bronfman}, {Henning}, {Kramer}, \&
  {Heitsch}}]{nguyenluong13}
{Nguyen Luong}, Q., {Motte}, F., {Carlhoff}, P., {et~al.} 2013, ArXiv e-prints

\bibitem[{{Nguyen Luong} {et~al.}(2011){Nguyen Luong}, {Motte}, {Hennemann},
  {Hill}, {Rygl}, {Schneider}, {Bontemps}, {Men'shchikov}, {Andr{\'e}},
  {Peretto}, {Anderson}, {Arzoumanian}, {Deharveng}, {Didelon}, {di Francesco},
  {Griffin}, {Kirk}, {K{\"o}nyves}, {Martin}, {Maury}, {Minier}, {Molinari},
  {Pestalozzi}, {Pezzuto}, {Reid}, {Roussel}, {Sauvage}, {Schuller}, {Testi},
  {Ward-Thompson}, {White}, \& {Zavagno}}]{nguyenluong11b}
{Nguyen Luong}, Q., {Motte}, F., {Hennemann}, M., {et~al.} 2011, \aap, 535, A76

\bibitem[{Padoan \& Nordlund(2004)}]{Padoan:2004ez}
Padoan, P. \& Nordlund, A. 2004, ApJ, 617, 559

\bibitem[{Padoan {et~al.}(1997)Padoan, Nordlund, \& Jones}]{Padoan:1997uu}
Padoan, P., Nordlund, A., \& Jones, B. J.~T. 1997, MNRAS, 288, 145

\bibitem[{Palau {et~al.}(2013)Palau, Estalella, Girart, \& et~al.}]{Palau:2013}
Palau, A., Estalella, R., Girart, J.~M., \& et~al. 2013, A{\&}A, accepted

\bibitem[{Palmeirim {et~al.}(2013)Palmeirim, Andr{\'e}, Kirk, Ward-Thompson,
  Arzoumanian, K{\"o}nyves, Didelon, Schneider, Benedettini, Bontemps,
  Di~Francesco, Elia, Griffin, Hennemann, Hill, Martin, Men'shchikov, Molinari,
  Motte, Nguyen-Luong, Nutter, Peretto, Pezzuto, Roy, Rygl, Spinoglio, \&
  White}]{Palmeirim:2013da}
Palmeirim, P., Andr{\'e}, P., Kirk, J., {et~al.} 2013, A{\&}A, 550, 38

\bibitem[{Passot \& V{\'a}zquez-Semadeni(1998)}]{Passot:1998cr}
Passot, T. \& V{\'a}zquez-Semadeni, E. 1998, Physical Review E (Statistical
  Physics, 58, 4501

\bibitem[{Peretto {et~al.}(2012)Peretto, Andr{\'e}, K{\"o}nyves, Schneider,
  Arzoumanian, Palmeirim, Didelon, Attard, Bernard, Di~Francesco, Elia,
  Hennemann, Hill, Kirk, Men'shchikov, Motte, Nguyen-Luong, Roussel, Sousbie,
  Testi, Ward-Thompson, White, \& Zavagno}]{Peretto:2012bu}
Peretto, N., Andr{\'e}, P., K{\"o}nyves, V., {et~al.} 2012, A{\&}A, 541, 63

\bibitem[{Peretto {et~al.}(2013)Peretto, Fuller, Duarte-Cabral, Avison,
  Hennebelle, Pineda, Andr{\'e}, Bontemps, Motte, Schneider, \&
  Molinari}]{Peretto:2013kt}
Peretto, N., Fuller, G.~A., Duarte-Cabral, A., {et~al.} 2013, A{\&}A, 555, 112

\bibitem[{Poulton {et~al.}(2008)Poulton, Robitaille, Greaves, Bonnell,
  Williams, \& Heyer}]{Poulton:2008fp}
Poulton, C.~J., Robitaille, T.~P., Greaves, J.~S., {et~al.} 2008, MNRAS, 384,
  1249

\bibitem[{Pound(1998)}]{Pound:1998hj}
Pound, M.~W. 1998, Astrophysical Journal Letters v.493, 493, L113

\bibitem[{Rivera-Ingraham {et~al.}(2013)Rivera-Ingraham, Martin, Polychroni,
  Motte, Schneider, Bontemps, Hennemann, Menshchikov, Nguyen~Luong, Andre,
  Arzoumanian, Bernard, Di~Francesco, Elia, Fallscheer, Hill, Li, Minier,
  Pezzuto, Roy, Rygl, Sadavoy, Spinoglio, White, \&
  Wilson}]{RiveraIngraham:2013wv}
Rivera-Ingraham, A., Martin, P.~G., Polychroni, D., {et~al.} 2013, arXiv, 1301,
  3805

\bibitem[{Roccatagliata {et~al.}(2013)Roccatagliata, Preibisch, Ratzka, \&
  Gaczkowski}]{Roccatagliata:2013wv}
Roccatagliata, V., Preibisch, T., Ratzka, T., \& Gaczkowski, B. 2013, arXiv,
  5201

\bibitem[{Rodgers {et~al.}(1960)Rodgers, Campbell, \&
  Whiteoak}]{Rodgers:1960tc}
Rodgers, A.~W., Campbell, C.~T., \& Whiteoak, J.~B. 1960, MNRAS, 121, 103

\bibitem[{Rom{\'a}n-Z{\'u}{\~n}iga \& Lada(2008)}]{RomanZuniga:2008tm}
Rom{\'a}n-Z{\'u}{\~n}iga, C.~G. \& Lada, E.~A. 2008, Handbook of Star Forming
  Regions, I, 928

\bibitem[{Russeil {et~al.}(2013)Russeil, Schneider, Anderson, \&
  et~al.}]{Russeil2013}
Russeil, D., Schneider, N., Anderson, L.~D., \& et~al. 2013, accepted in A\&A

\bibitem[{Schneider {et~al.}(2013)Schneider, Andr\'e, K\"onyves, \&
  et~al.}]{schneider2013}
Schneider, N., Andr\'e, P., K\"onyves, V., \& et~al. 2013, accepted in ApJ

\bibitem[{Schneider {et~al.}(2010{\natexlab{a}})Schneider, Csengeri, Bontemps,
  Motte, Simon, Hennebelle, Federrath, \& Klessen}]{Schneider:2010kx}
Schneider, N., Csengeri, T., Bontemps, S., {et~al.} 2010{\natexlab{a}}, A{\&}A,
  520, 49

\bibitem[{Schneider {et~al.}(2012{\natexlab{a}})Schneider, Csengeri, Hennemann,
  Motte, Didelon, Federrath, Bontemps, Di~Francesco, Arzoumanian, Minier,
  Andr{\'e}, Hill, Zavagno, Nguyen-Luong, Attard, Bernard, Elia, Fallscheer,
  Griffin, Kirk, Klessen, K{\"o}nyves, Martin, Men'shchikov, Palmeirim,
  Peretto, Pestalozzi, Russeil, Sadavoy, Sousbie, Testi, Tremblin,
  Ward-Thompson, \& White}]{Schneider:2012ds}
Schneider, N., Csengeri, T., Hennemann, M., {et~al.} 2012{\natexlab{a}},
  A{\&}A, 540, L11

\bibitem[{Schneider {et~al.}(2012{\natexlab{b}})Schneider, G{\"u}sten,
  Tremblin, Hennemann, Minier, Hill, Comer{\'o}n, Requena-Torres, Kraemer,
  Simon, R{\"o}llig, Stutzki, Djupvik, Zinnecker, Marston, Csengeri, Cormier,
  Lebouteiller, Audit, Motte, Bontemps, Sandell, Allen, Megeath, \&
  Gutermuth}]{Schneider:2012hz}
Schneider, N., G{\"u}sten, R., Tremblin, P., {et~al.} 2012{\natexlab{b}},
  A{\&}A, 542, L18

\bibitem[{Schneider {et~al.}(2010{\natexlab{b}})Schneider, Motte, Bontemps,
  Hennemann, Di~Francesco, Andr{\'e}, Zavagno, Csengeri, Men'shchikov, Abergel,
  Baluteau, Bernard, Cox, Didelon, di~Giorgio, Gastaud, Griffin, Hargrave,
  Hill, Huang, Kirk, K{\"o}nyves, Leeks, Li, Marston, Martin, Minier, Molinari,
  Olofsson, Panuzzo, Persi, Pezzuto, Roussel, Russeil, Sadavoy, Saraceno,
  Sauvage, Sibthorpe, Spinoglio, Testi, Teyssier, Vavrek, Ward-Thompson, White,
  Wilson, \& Woodcraft}]{Schneider:2010ec}
Schneider, N., Motte, F., Bontemps, S., {et~al.} 2010{\natexlab{b}}, A{\&}A,
  518, L83

\bibitem[{Schneider {et~al.}(1998{\natexlab{a}})Schneider, Stutzki,
  Winnewisser, \& Block}]{Schneider:1998ta}
Schneider, N., Stutzki, J., Winnewisser, G., \& Block, D. 1998{\natexlab{a}},
  A{\&}A, 335, 1049

\bibitem[{Schneider {et~al.}(1998{\natexlab{b}})Schneider, Stutzki,
  Winnewisser, Poglitsch, \& Madden}]{Schneider:1998vx}
Schneider, N., Stutzki, J., Winnewisser, G., Poglitsch, A., \& Madden, S.
  1998{\natexlab{b}}, A{\&}A, 338, 262

\bibitem[{Thompson {et~al.}(2012)Thompson, Urquhart, Moore, \&
  Morgan}]{Thompson:2012gn}
Thompson, M.~A., Urquhart, J.~S., Moore, T. J.~T., \& Morgan, L.~K. 2012,
  MNRAS, 421, 408

\bibitem[{Tremblin {et~al.}(2012{\natexlab{a}})Tremblin, Audit, Minier, \&
  {\SortNoop{a}}{S}chneider}]{Tremblin:2012ej}
Tremblin, P., Audit, E., Minier, V., \& {\SortNoop{a}}{S}chneider, N.
  2012{\natexlab{a}}, A{\&}A, 538, 31

\bibitem[{Tremblin {et~al.}(2012{\natexlab{b}})Tremblin, Audit, Minier,
  Schmidt, \& Schneider}]{Tremblin:2012he}
Tremblin, P., Audit, E., Minier, V., Schmidt, W., \& Schneider, N.
  2012{\natexlab{b}}, A{\&}A, 546, 33

\bibitem[{Urquhart {et~al.}(2003)Urquhart, White, Pilbratt, \&
  Fridlund}]{Urquhart:2003jj}
Urquhart, J.~S., White, G.~J., Pilbratt, G.~L., \& Fridlund, C. V.~M. 2003,
  A{\&}A, 409, 193

\bibitem[{V{\'a}zquez-Semadeni(1994)}]{VazquezSemadeni:1994fm}
V{\'a}zquez-Semadeni, E. 1994, Astrophysical Journal v.423, 423, 681

\bibitem[{V{\'a}zquez-Semadeni {et~al.}(2008)V{\'a}zquez-Semadeni,
  Gonz{\'a}lez, Ballesteros-Paredes, Gazol, \& Kim}]{VazquezSemadeni:2008ho}
V{\'a}zquez-Semadeni, E., Gonz{\'a}lez, R.~F., Ballesteros-Paredes, J., Gazol,
  A., \& Kim, J. 2008, MNRAS, 390, 769

\bibitem[{Verma {et~al.}(1994)Verma, Bisht, Ghosh, Iyengar, Rengarajan, \&
  Tandon}]{Verma:1994vt}
Verma, R.~P., Bisht, R.~S., Ghosh, S.~K., {et~al.} 1994, A{\&}A, 284, 936

\bibitem[{White {et~al.}(1999)White, Nelson, Holland, Robson, Greaves,
  McCaughrean, Pilbratt, Balser, Oka, Sakamoto, Hasegawa, McCutcheon, Matthews,
  Fridlund, Tothill, Huldtgren, \& Deane}]{White:1999ue}
White, G., Nelson, R.~P., Holland, W.~S., {et~al.} 1999, A{\&}A, 342, 233

\bibitem[{Whitworth(1979)}]{Whitworth:1979uc}
Whitworth, A.~P. 1979, MNRAS, 186, 59

\bibitem[{Williams {et~al.}(1994)Williams, de~Geus, \&
  Blitz}]{Williams:1994hla}
Williams, J.~P., de~Geus, E.~J., \& Blitz, L. 1994, ApJ, 428, 693

\bibitem[{Zavagno {et~al.}(2007)Zavagno, Pomar{\`e}s, Deharveng, Hosokawa,
  Russeil, \& Caplan}]{Zavagno:2007ix}
Zavagno, A., Pomar{\`e}s, M., Deharveng, L., {et~al.} 2007, A{\&}A, 472, 835

\bibitem[{Zavagno {et~al.}(2010)Zavagno, Russeil, Motte, Anderson, Deharveng,
  Rod{\'o}n, Bontemps, Abergel, Baluteau, Sauvage, Andr{\'e}, Hill, \&
  White}]{Zavagno:2010jv}
Zavagno, A., Russeil, D., Motte, F., {et~al.} 2010, A{\&}A, 518, L81

\end{thebibliography}

\end {document}